\documentclass[11pt,reqno]{article}
\usepackage{bm}
\usepackage{amsthm}
\usepackage{amsmath}
\usepackage{amssymb}
\usepackage[round]{natbib}
\usepackage{color}
\usepackage{authblk}
\usepackage{microtype}
\usepackage{graphicx}
\usepackage{float}
\usepackage{prodint}
\usepackage{paralist}
\usepackage{epstopdf}
\usepackage{geometry}
\DeclareMathSymbol{\shortminus}{\mathbin}{AMSa}{"39}
\newcommand{\rnc}{\renewcommand}
\newcommand{\nc}{\newcommand}
\newcommand{\mrm}{\mathrm}

\nc{\mb}{\mathbb}
\nc{\mc}{\mathcal}
\nc{\E}{\mb{E}}
\nc{\N}{\mb{N}}
\nc{\R}{\mb{R}}
\nc{\Q}{\mb{Q}}
\rnc{\P}{\mrm P}
\rnc{\d}{\mrm d}
\nc{\C}{\mc{C}}
\nc{\D}{\mc{D}}
\nc{\B}{\mc{B}}
\nc{\vbeta}{\bm \beta}
\nc{\vtheta}{\bm \theta}
\nc{\vX}{\bm X}
\nc{\vy}{\bm y}
\nc{\vU}{\bm U}
\nc{\vI}{\bm I}
\nc{\vE}{\bm E}
\nc{\ve}{\bm e}
\nc{\vV}{\bm V}
\nc{\vv}{\bm v}
\nc{\vS}{\bm S}
\nc{\vSigma}{\bm \Sigma}
\nc{\oPo}{\stackrel{\mrm p}{\rightarrow}}
\nc{\oWo}{\stackrel{w}{\rightarrow}}
\nc{\oDo}{\stackrel{d}{\longrightarrow}}
\nc{\eff}{\|F\|}

\def\E{{ E }}
\def\R{{ \mathbb{R} }}
\def\N{{ \mathbb{N} }}

\def\P{ P }

\def\E{ E }

\newtheorem{lemma}{Lemma}
\newtheorem{assump}{Assumption}

\newtheorem{theorem}{Theorem}
\newtheorem{prop}{Proposition}

\newcommand\blfootnote[1]{%
  \begingroup
  \renewcommand\thefootnote{}\footnote{#1}%
  \addtocounter{footnote}{-1}%
  \endgroup
} 

\begin{document}

\title{\Large \bf Inferring median survival differences in general factorial designs via
	permutation tests}
\author[1,$*$]{Marc Ditzhaus}
\author[2]{Dennis Dobler}
\author[1]{Markus Pauly}

\affil[1]{Department of Statistics, TU Dortmund University, Germany.}
\affil[2]{Department of Mathematics, Vrije Universiteit Amsterdam, The Netherlands.}

\maketitle

\begin{abstract}
\blfootnote{${}^*$ e-mail: marc.ditzhaus@tu-dortmund.de}
 Factorial survival designs with right-censored observations are commonly inferred by Cox regression and explained by means of hazard ratios. However, in case of non-proportional hazards, their interpretation can become cumbersome; especially for clinicians. We therefore offer an alternative: median survival times are used to estimate treatment and interaction effects and null hypotheses are formulated in contrasts of their population versions. Permutation-based tests and confidence regions are proposed and shown to be asymptotically valid. Their type-1 error control and power behavior are investigated in extensive simulations, showing the new methods' wide applicability. The latter is complemented by an illustrative data analysis.
\end{abstract}



\section{Introduction}\label{sec:intro}
Medical studies with time-to-event endpoints are most commonly inferred by means of logrank tests or Cox regression models. In case of non-proportional hazards, however, they are not always the most appropriate choice and several alternatives to logrank tests have been proposed  \citep{brendel2014weighted, ditzhaus:pauly:2019, ditzhaus:friedrich:2018, ditzhaus2020permutation, gorfine2020}. 
In factorial survival designs this issue is additionally hampered by the desire to 
summarize main treatment and interaction effects in single quantities. 
For hazard ratios as effect estimates, this can only be achieved under the proportional hazards assumption. As this can be hard to verify or simply wrong, some alternatives to hazard ratios as effect sizes are desired that provide `{\it clinically meaningful outcomes for (...) patients}' and for `{\it discussing the value of clinical trials with patients}' \citep{ellis2014american}. Proposed alternatives cover average hazard ratios \citep{kalbfleisch1981estimation, bruckner2017sequential}, concordance and Mann-Whitney effects \citep{koziol2009concordance, dopa2018, dobler:pauly:2019} the (restricted) mean survival \citep{royston2013restricted, ben2019median} or median survival times \citep{brookmeyer1982confidence,brookmeyerCrowley1982,chenZhang2016}. 

Except for Mann-Whitney effects and concordance odds \citep{martinussen2013estimation, dobler:pauly:2019}, statistical tools for inferring these quantities have mostly been developed for special two or multiple sample settings. Applying these methods to factorial designs with at least two crossed factors, would neither use the factorial structure to full capacity nor allow for the quantification of potential interaction effects.
But this is one of the key reasons to plan studies with a factorial design. 

It is the aim of the present paper to increase the urgently needed flexibility to apply appropriate procedures that take full advantage of factorial structures in the analysis of time-to-event data \citep{green2005factorial}. 
As the handling of the above-mentioned effects sizes requires different approaches of diverse complexity, it is impossible to treat all of them simultaneously. We thus focus on the median survival times as effect estimates and leave factorial extensions of the others for future research. 

Median survival times are `{\it easy to understand}' \citep{ben2019median} and therefore frequently reported in medical studies. In fact, they are `{\it the most common measure used in the outcome reporting of oncology clinical trials}' \citep{ben2019median} and their application has recently been propagated \citep{chenZhang2016}. 
Besides their use as a descriptive tool, they are also applied for inferential conclusions. The methodological foundation for the latter was laid in the works of \cite{brookmeyer1982confidence,brookmeyerCrowley1982} and later extended by others 
 \citep{su1993nonparametric,chen2007covariates,chenZhang2016}. All these investigations focus on specific aspects of two or $k$ sample settings and do not cover potential interactions. 
However, for non-censored observations, extensions to general factorial designs 
have recently been proposed \citep{ditzhaus2019qanova}. 
These are based on the idea of using heterogeneity robust studentized permutation tests \citep{janssenPauls2003,chungRomano2013,paulyETAL2015} in statistics of Wald-type. Thus, it appeared  natural to transfer these ideas to time-to-event situations. It thereby turned out that censoring makes theoretical investigations much more demanding. Moreover, the assumption of a general censoring mechanism, that can be arbitrary across groups, additionally hampers the construction of valid estimators. All this made the study of the robust permutation approach even more difficult. Carefully extending some results on empirical processes we nevertheless accomplished this which finally resulted in asymptotically valid and consistent studentized permutation tests for general factorial survival designs. 
These tests are even finitely exact in the special case of exchangeable data.

The paper is organized as follows.
The model and all mathematical notations are introduced in 
Section~\ref{sec:setup}.
Section~\ref{sec:asy_results} treats the construction and the asymptotic properties of a sensible test statistic.
Its permutation version is introduced and analyzed in Section~\ref{sec:perm}.
Section~\ref{sec:sim} offers the results of extensive simulation studies in which the proposed tests are compared with a competitor from quantile regression. 
A real data analysis in Section~\ref{sec:real_data} focuses on data from trials on liver diseases. We conclude the paper with a discussion in Section~\ref{sec:discuss}.
All proofs are given in the appendix.

\section{Factorial survival set-up}
\label{sec:setup}

To simplify notation, we regard any factorial design for now as a $k$-sample set-up:
\begin{align}\label{eqn:model}
	T_{ij}\sim S_i,\quad C_{ij}\sim G_{i}, \quad i=1,\ldots,k, \ \  j=1,\ldots,n_i,
\end{align}
where $T_{ij}$ and $C_{ij}$ respectively denote the 
independent survival and censoring times of individual $j$ in group $i$. 
The survival functions $S_i$ are assumed to be continuous with density functions $f_i$; 
the censoring survival functions $G_i$ need not be continuous.
The actually observable data consist of the right-censored event times $X_{ij}=\min(T_{ij},C_{ij})$ and the corresponding censoring statuses $\delta_{ij}=1\{X_{ij}=T_{ij}\}$.
In this paper the quantities of main interest are the group-specific (survival) medians 
\begin{align}\label{eqn:def_quantiles}
	m_i=S_i^{-1}(1/2) = \inf\{ t\in \R : S_i(t)\leq 1/2\},  \quad i=1,\ldots,k.
\end{align}
We wish to test for relationships of the medians in various factorial designs:
\begin{align*}
	\mathcal H_0(\mathbf{H}): \mathbf{H} \mathbf{m} = \mathbf{0},
\end{align*}
where $\mathbf{m}=(m_1,\dots,m_k)'\in \R^k$, \ $\mathbf{0}\in\R^l$ is the zero vector and $\mathbf{H}\in \R^{l\times k}$ denotes a contrast matrix, i.e. the rows sum up to $0$. For example, the centering matrix $\mathbf{H} = \mathbf{P}_k =  \mathbf I_k - \mathbf{J}_k/k$ is used to describe the null hypothesis $\mathcal H_0(\mathbf{P}_k): m_{1}=\ldots=m_{k}$ of no group effects in the $k$-sample setting, where $\mathbf{I}_k$ is the $k\times k$-dimensional unit matrix and $\mathbf{J}_k\in \R^{k \times k}$ consists of $1$'s only.\\ 
Two-way designs are incorporated by splitting up the indices $ i = (i_A,i_B)$ for factors $A$ (with $a \in \N$ levels) and B (with $b \in \N$ levels). 
Then, the contrast matrices for testing hypotheses about no main or interaction effects are given by
\begin{itemize}
	\item $\mathbf{H}_{A}= \mathbf{P}_a \otimes (\mathbf{J}_b/b)$ with \ $\mathcal H_0(\mathbf{H}_{A}) = \{\mathbf{H}_{A}\mathbf{m} = \mathbf{0} \} = \{ \bar{m}_{1\cdot} = \ldots = \bar{m}_{a\cdot}\}$ 
	(\textit{no main effect of factor A}),
	
	\item $\mathbf{H}_B = (\mathbf{J}_a/a) \otimes \mathbf{P}_b$ with \ $\mathcal H_0(\mathbf{H}_{B}) = \{\mathbf{H}_{B}\mathbf{m} = \mathbf{0} \} = \{ \bar{m}_{\cdot1} = \ldots = \bar{m}_{\cdot b}\}$ (\textit{no main effect of factor B}),
	
	\item $\mathbf{H}_{AB} = \mathbf{P}_a \otimes \mathbf{P}_b$ with \ $\mathcal H_0(\mathbf{H}_{AB}) = \{\mathbf{H}_{AB}\mathbf{m} = \mathbf{0} \} = \{ \bar{m}_{i_1i_2} - \bar{m}_{\cdot i_2} - \bar{m}_{i_1\cdot} + \bar{m}_{\cdot \cdot} = 0 \ \text{for all } i_1, i_2\}  $ \ (\textit{no interaction effect}).
\end{itemize}
Here, $\otimes$ denotes the Kronecker product and $\bar{m}_{i_1 \cdot}$, $\bar{m}_{\cdot i_2}$ and $\bar{m}_{\cdot \cdot}$ are the means over the dotted indices. Extensions to higher-way crossed or hierachically nested layouts can be obtained similarly \citep{paulyETAL2015,ditzhaus2019qanova}.  

Estimates of the survival medians will be based on the Kaplan--Meier estimators given by
\begin{align*}
	\widehat S_i(t)=\prod_{j:X_{ij}\leq t}\Bigl( 1- \frac{\delta_{ij}}{Y_i(X_{ij})} \Bigr), \quad t\geq 0, \quad i=1,\ldots,k.
\end{align*}
Here, $Y_i(t)=\sum_{j=1}^{n_i} 1\{X_{ij}\geq t\}$ is the number of subjects in group $i$ being at risk just before $t$. Then the natural estimator of the median $m_i$ is 
\begin{align}\label{eqn:def_qhat}
	\widehat m_i = \widehat S_i^{-1}(1/2) = \inf\{ t\in \R : \widehat S_i(t)\leq 1/2\},  \quad i=1,\ldots,k.
\end{align}	     
In the same way, $\widehat S_i^{-1}(q)$ is an estimator of the $q$-th quantile $S_i^{-1}(q)$, $q\in(0,1)$. 

\section{The asymptotic Wald-type tests}\label{sec:asy_results}

Inference on the population medians will be achieved by means of Wald-type tests.
To this end, let us first introduce the so-called projection matrix \citep{brunner:dette:munk:1997,paulyETAL2015,smaga2017diagonal} $\mathbf{T}=\mathbf{H}'(\mathbf{HH}')^+ \mathbf{H}$, where $(\cdot)^+$ denotes the Moore--Penrose inverse. 
The benefits of $\mathbf{T}$ are its symmetry and idempotence.
The null hypotheses are unaffected by this change of contrast matrices \citep{brunner:dette:munk:1997}: 
$\mathbf{T}\mathbf{m}=\textbf{0} $ holds if and only if $ \mathbf{Hm}=\mathbf{0}$. 
Then a Wald-type test statistic for this null hypothesis is given as
\begin{align*}
	W_n(\mathbf{T}) = n( \mathbf{T} \widehat{\mathbf{m}})' ( \mathbf{T} \widehat{\mathbf{\Sigma}} \mathbf{T}' )^{+} \mathbf{T} \widehat{\mathbf{m}}\quad \text{with} \quad \widehat{\mathbf{\Sigma}} = \text{diag} \Big(\frac n {n_1} \widehat\sigma_1^2,\ldots, \frac n {n_k}  \widehat \sigma_k^2 \Big).
\end{align*}
Here, $n= n_1 + \dots + n_k$ is the total sample size and
$\widehat{\mathbf{m}}=(\widehat m_1,\dots,\widehat m_k)'$ is the vector of pooled median estimators. Moreover, $\widehat \sigma_i^2$ denote consistent estimators of the asymptotic variances $\sigma_i^2$ of $\sqrt{n_i}(\widehat m_i - m_i)$ which are yet to be determined. In fact, discussing the weak convergence of $W_n(\mathbf{T})$ 
boils down to investigating the convergence of the median and subsequent variance estimators.
To this end denote by $\stackrel d \longrightarrow$ convergence in distribution and 
by $\stackrel p \longrightarrow$ convergence in probability.
Throughout the paper we make the following assumption which ensures the existence of $\widehat m_i = \widehat S_i^{-1}(1/2)$ with a probability tending to 1
and the asymptotic normality of $\widehat m_i$ as $n_i \rightarrow \infty$.
\begin{assump}\label{ass:densities} For each sample group $i=1,\dots,k$ it holds that
	\phantom{X}
	\begin{itemize}
		\item[\textnormal{(a)}] The density is positive at the median: $f_i(m_i)>0$.
		\item[\textnormal{(b)}]\label{enu:ass:densities:G>0} It is possible to observe events after the median survival time: $G_i(m_i) > 0$.
	\end{itemize}	 
\end{assump} 

A variant of the following proposition traces back to \cite{sander:1975}.
\begin{prop}\label{PROP:UNCOND}
	Suppose Assumption~\ref{ass:densities}. As $n_i\rightarrow \infty$, 
	$
	\sqrt{n_i}( \widehat m_i -  m_i) \stackrel d \longrightarrow Z_i \sim \mathcal{N}(0, \sigma_i^2)
	$ 
	with variance 
	\begin{align}\label{eqn:def_sigmai}
		\sigma^2_i = \frac{-1}{4f_i(m_i)^{2}}  \int_0^{m_i} \frac{\mathrm{ d }S_i}{G_{i} S_i^2}.
	\end{align} 
\end{prop}

Proposition~\ref{PROP:UNCOND} shows that estimators of the asymptotic variances should involve an estimate of $f_i(m_i)$. For computational reasons we propose to use an estimator of the standard deviation which, in the uncensored case, is based on a standardized confidence interval \citep{mckeanSchrader1984}; 
see \cite{priceBonett2001} for a slight modification to improve the estimator's performance for small sample sizes. \cite{chakraborti:1988,chakraborti:1990} extended the estimator to the censored case and proved its consistency. For our purposes we use different variants of this estimator. The first is given by
\begin{align}\label{eqn:schrader+mckean_est}
	\widehat \sigma_{i,\text{two-sided}} = 0.5 \ z_{\gamma/2}^{-1} \  \sqrt{n}_i(\widehat S_i^{-1}(l_i) - \widehat S_i^{-1}(u_i)),
\end{align}
where $\gamma\in(0,1)$ is a fixed number, $z_{\alpha}= \Phi^{-1}(1-\alpha), \alpha \in (0,1)$, is the $(1-\alpha)$-quantile of the standard normal distribution function $\Phi$ and we set
\begin{align*}
	&l_i = 0 \vee 0.5 \Bigl( 1  - z_{\gamma/2} \sqrt{\widehat V_i/n_i} \Bigr) \quad \text{and}  \quad  u_i = 1 \wedge 0.5 \Bigl(1 + z_{\gamma/2} \sqrt{\widehat V_i/n_i} \Bigr)
	\quad \text{for} \quad
	\widehat V_i = n_i \sum_{j: X_{ij} \leq \widehat m_i} \frac{\delta_{ij}}{Y_i^2(X_{ij})}.
\end{align*}
Here, $c \vee d$ and $c \wedge d$ denote the maximum and minimum of two real numbers $c$ and $d$, respectively. We note that $\widehat V_i$ is a consistent estimator of the asymptotic variance of the normalized Nelson--Aalen estimator at $m_i$ \citep{aalen1976}.

In cases of strong censoring in combination with small sample sizes $\widehat S_i^{-1}(l_i)$ may not exist. Here, one solution is to adjust the involved standard normal quantile; see also 
\cite{priceBonett2001} for a short discussion on the choice of $\gamma$.  Usually, $\gamma =5\%$ or $10\%$ is chosen. 
If $\widehat S_i^{-1}(l_i)$ does not exist for the previously made choice of $\gamma$, we suggest to use the Kaplan-Meier estimator at the last observed event time $\tilde l_i := \min_{t > 0} \widehat S_i(t)$ instead of $l_i$. In this case we solve for $\gamma$ in the original definition of $l_i$:
$$ \tilde \gamma :=  2 \Phi \Big( (1 - 2 \tilde l_i )/ \sqrt{\widehat{V}_i /n_i} ) \Big). $$
This $\tilde \gamma$ then gives us $\tilde u_i =
1 \wedge 0.5 (1 + z_{\tilde{\gamma}/2} ({\widehat V_i/n_i})^{1/2})$ instead of $u_i$. 
Finally, the adjusted estimator of the standard deviation is obtained according to Formula~\eqref{eqn:schrader+mckean_est} with $\tilde \gamma$, $\tilde l_i$, and $\tilde u_i$ instead of $\gamma$, $l_i$, and $u_i$, respectively.
Note that this adjustment does not play a role as $n_i \to \infty$ 
because $\widehat S_i^{-1}(l_i)$ exists with a probability tending to 1 for any choice of $\gamma$.

A different approach to overcome the problem of the non-existence $\widehat S_i^{-1}(l_i)$ is to switch from two-sided intervals to one-sided intervals \citep{chakraborti:1988}. In this case, the estimator is given by
\begin{align}\label{eqn:sigma_onesided}
	\widehat \sigma_{i,\text{one-sided}} = z_{\gamma/2}^{-1} \  \sqrt{n}_i(\widehat m_i - \widehat S_i^{-1}(u_i)).
\end{align}

To unify notation, we subsequently suppose that $\widehat \sigma_i$ is either the adjusted two-sided interval variance estimator from \eqref{eqn:schrader+mckean_est} or the one-sided counterpart from \eqref{eqn:sigma_onesided}. While the theory developed below is valid for both choices, the simulations in Section~\ref{sec:sim} show a slightly advantageous power behavior for the one-sided version.
\begin{lemma}\label{LEM:EST_F_CONSIS}
	As $n_i \to \infty$, we have $\widehat \sigma_i \overset{p}{\longrightarrow} \sigma_i$ for each $i=1,\dots,k$.
\end{lemma}

By combining all of the convergences discussed above,
we wish to find the limit null distribution of $W_n(\mathbf{T})$.
However, as the sample sizes might grow at different paces, we need to make the following weak assumption that, asymptotically, no sample group vanishes in relation to any other group:
\begin{assump}
	\label{ass:liminf}
	$\min\limits_{i=1,\dots, k}\liminf\limits_{n\to \infty} \frac{n_i} n > 0.$ 
\end{assump}
Under this assumption it follows that the vector $\widehat{\textbf{m}}$ of estimated medians converges in distribution at $\sqrt{n}$-rate. In particular, we obtain the following theorems:

\begin{theorem}\label{THEO:TESTSTAT_UNCON}
	Suppose that Assumptions~\ref{ass:densities} and~\ref{ass:liminf} hold. then, we have
	\begin{itemize}
		\item[\textnormal{(a)}]
		under $\mathcal H_0(\mathbf{T}): \mathbf{T}\mathbf{m} = \mathbf{0}$,  \  $W_n(\mathbf{T}) \stackrel d \longrightarrow Z \sim\chi^2_{\text{rank}(\mathbf{T})}$ as $n\rightarrow \infty$.
		\item[\textnormal{(b)}] 
		under $\mathcal H_1(\mathbf{T}): \mathbf{T}\mathbf{m} \neq \mathbf{0}$, \ $W_n(\mathbf{T}) \stackrel p \longrightarrow\infty$  as $n\rightarrow \infty$.
	\end{itemize}
\end{theorem}

For $\alpha \in (0,1)$ denote by $\chi^2_{\ell; \alpha}$ the $(1-\alpha)$-quantile of a $\chi^2$-distribution with $\ell \in \N$ degrees of freedom. 
\newtheorem{cor}{Corollary}
\begin{cor}\label{cor:test_asy}
	Let $\alpha \in (0,1)$ and suppose that Assumptions~\ref{ass:densities} and~\ref{ass:liminf} hold.
	The asymptotic test $\varphi_n = 1\{ W_n(\mathbf{T}) > \chi^2_{\text{rank}(\mathbf{T}); \alpha} \}$ for $\mathcal{H}_0(\mathbf{T})$ versus $\mathcal{H}_1(\mathbf{T})$ is consistent with asymptotic level $\alpha$, i.e. as $n \to \infty$,
	$$ E(\varphi_n) = P(W_n(\mathbf{T}) > \chi^2_{\text{rank}(\mathbf{T}); \alpha}) \longrightarrow 1_{\mathcal{H}_1(\mathbf{T})} + \alpha \cdot 1_{\mathcal{H}_0(\mathbf{T})}. $$
\end{cor}

\section{The permutation test}\label{sec:perm}

Theorem~\ref{THEO:TESTSTAT_UNCON} is already sufficient to deduce consistent tests for $\mathcal{H}_0(\mathbf{T})$ versus $\mathcal{H}_1(\mathbf{T})$.
However, if the sample sizes $n_1, \dots, n_k$ are not large enough, the type-1 error rates of the tests based on Wald-type statistics are often inflated \citep{paulyETAL2015,ditzhaus2019qanova}; see also Section~\ref{sec:sim} below.
To solve this problem, we suggest to conduct the new Wald-type test as a robust permutation test.
Denote by $\pi = (\pi(1), \pi(2), \dots, \pi(n))$ a random vector that is uniformly distributed on the set of all permutations of $(1, \dots, n)$.
Writing the pooled data as 
$$ \mathbf{X} = (X_1, \delta_1, X_2, \delta_2, \dots, X_n, \delta_n) = (X_{11}, \delta_{11}, X_{12}, \delta_{12}, \dots, X_{k, n_k}, \delta_{k,n_k}), $$
the permuted new groups are given as 
$$ \mathbf{X}_1^\pi = (X_{\pi(1)}, \delta_{\pi(1)},  \dots, X_{\pi(n_1)}, \delta_{\pi(n_1)}) \ ,  \quad 
\dots \ , \quad \mathbf{X}_k^\pi = (X_{\pi(n-n_k+1)}, \delta_{\pi(n-n_k+1)}, \dots, X_{\pi(n)}, \delta_{\pi(n)}). $$
We write $\widehat S_i^\pi$ and $\widehat \sigma_i^\pi$ for the Kaplan-Meier and standard deviation estimator from Section~\ref{sec:asy_results} based on $\mathbf{X}_i^\pi$, $i=1,\dots, k$, respectively. Similarly, each $\widehat m^\pi_i$ is defined in terms of $\widehat S_i^\pi$.
Finally, the permutation Wald-type statistic results as 
$$ W_n^\pi(\mathbf{T}) = n( \mathbf{T} \widehat{\mathbf{m}}^\pi)' ( \mathbf{T} \widehat{\mathbf{\Sigma}}^\pi \mathbf{T}' )^{+} \mathbf{T} \widehat{\mathbf{m}}^\pi, $$
where $\widehat{\mathbf{m}}^\pi = (\widehat m^\pi_1, \dots, \widehat m^\pi_k)'$ and $\widehat{\mathbf{\Sigma}}^\pi = \text{diag}(\frac n {n_1}\widehat \sigma_1^{\pi2},\ldots,\frac n {n_k} \widehat \sigma_k^{\pi2} )$.
Theoretical analyses from the appendix show that the permutation medians $\widehat m^\pi_i$ behave similar to the pooled median estimator $\widehat m = \widehat S^{-1}(0.5)$ where $\widehat S$ denotes the pooled Kaplan-Meier estimator. 
That is, $\widehat S$ uses the complete dataset $\mathbf{X}$. 
Denote by $S$ the limit in probability of $\widehat S$; the convergence is argued in the appendix where also an explicit formula for $S$ is given.
For the existence and the convergence of the pooled median estimator and the permutation medians, 
we make the following assumption on the censoring distributions:
\begin{assump}
	\label{ass:pooled}
	\phantom{X} { }
	\begin{itemize}
		\item[\textnormal{(a)}] The function $f_0 := \sum_{i=1}^k f_i$ is positive at  $m = S^{-1}(0.5)$.
		\item[\textnormal{(b)}]  
		The pooled median is observable, i.e.\ $\min_{i=1,\dots,k} G_i(m) > 0$.
		\item[\textnormal{(c)}] The functions $f_i$ and $G_i$, $i=1,\dots, k$, are continuous on a neighborhood of $m$.
	\end{itemize}
\end{assump}
\noindent 
Under these assumptions, the studentized permutation approach works as stated below.

\begin{theorem}\label{THEO:STAT_PERM}
	Suppose that Assumptions~\ref{ass:densities}--\ref{ass:pooled} hold.
	Under $\mathcal H_0(\mathbf{T}):\mathbf{T}\mathbf{m}=\mathbf{0}$ and also under $\mathcal H_1(\mathbf{T}):\mathbf{T}\mathbf{m}\neq\mathbf{0}$, the conditional distribution of the permutation version $W_n^\pi(\mathbf{T})$ of $W_n(\mathbf{T}) $ given $\mathbf{X}$ always approaches the null distribution of $W_n(\mathbf{T})$, i.e.\ as $n \to \infty$,
	\begin{align*}
		\sup_{x\in\R} \Bigl |  \P \Bigl( W_n^\pi(\mathbf{T}) \leq x \  \vert \  \mathbf{X} \Bigr) - \P_{\mathcal{H}_0(\textbf{T})} \Bigl( W_n(\mathbf{T}) \leq x  \Bigr)  \Bigr| \stackrel p \longrightarrow 0.
	\end{align*}
\end{theorem}
\noindent
This result enables the construction of a consistent permutation test for $\mathcal H_0(\mathbf{T}):\mathbf{T}\mathbf{m}=\mathbf{0}$ versus $\mathcal H_1(\mathbf{T}):\mathbf{T}\mathbf{m}\neq\mathbf{0}$:
denote by $c^\pi_{\alpha}$ the $(1-\alpha)$-quantile of the conditional distribution of $W_n^\pi(\mathbf{T})$ given $\mathbf{X}$.
Then Lemma 1 and Theorem 7 in \cite{janssenPauls2003} ensure the convergence of $c^\pi_{\alpha}$ to $\chi^2_{rank(\mathbf{T});\alpha}$ in probability. This yields

\begin{cor}\label{cor:test_perm}
	Let $\alpha \in (0,1)$ and suppose that Assumptions~\ref{ass:densities}--\ref{ass:pooled}  hold.
	The permutation test $\varphi^\pi_n = 1\{ W_n(\mathbf{T}) > c^\pi_{\alpha} \}$ for $\mathcal{H}_0(\mathbf{T})$ versus $\mathcal{H}_1(\mathbf{T})$ is consistent and has asymptotic level $\alpha$, i.e.\ as $n \to \infty$,
	$$ E(\varphi^\pi_n) = P(W_n(\mathbf{T}) > c^\pi_{\alpha}) \longrightarrow 1_{\mathcal{H}_1(\mathbf{T})} + \alpha \cdot 1_{\mathcal{H}_0(\mathbf{T})}. $$
\end{cor}
Beyond this asymptotic correctness for general models, the permutation test has yet another beneficial property: in the special case that all sample groups share the same underlying distributions, i.e.\ if the observed data pairs are exchangeable, a randomized version of the test can be shown to be finitely exact, for all possible sample size combinations. We refer to \cite{hemerik2018} and the references cited therein for further reading on the exactness of permutation tests. Typically, permutation tests such as $\varphi^\pi_n$ also show a good type-1 error rate control for non-exchangeable situations in the case of small samples. We will empirically assess this finite sample behaviour in the subsequent Section~\ref{sec:sim}.

\section{Simulations}\label{sec:sim}
To complement our asymptotic results, we conducted an extensive simulation study to cover various small sample size settings. 
For ease of presentation, we focused on the following two-factorial {\bf designs.}
\begin{itemize}
	\item[(a)] A \textbf{$\mathbf{2\times 2}$ layout} with factors $A$ and $B$ (each with $a=b=2$ levels) in which we test for the presence of a main effect of factor $A$ and of an interaction effect by choosing the contrast matrices $\mathbf{H}=\mathbf{H}_{A}$ and $\mathbf{H}=\mathbf{H}_{AB}$, respectively. Here, we considered two different sample size scenarios $\mathbf{n}_1=(n_{11},n_{12},n_{21},n_{22})=(12,12,12,12)$ (\textit{balanced}) and $\mathbf{n}_2 = (16,11,7,14)$ (\textit{unbalanced}), and three different censoring settings with censoring rate vectors $\mathbf{cr}_1=(cr_{11},cr_{12},cr_{21},cr_{22})=(0.07,0.12,0.12,0.07)$ (\textit{low censoring}), $\mathbf{cr}_2=(0.29,0.38,0.25,0.35)$ (\textit{high censoring}) and $\mathbf{cr}_3=(0.12,0.38,0.07,0.29)$ (\textit{low/high censoring}).
	
	\item[(b)] A \textbf{$\mathbf{2\times 3}$ layout} with factors $A$ (possessing $a=2$ levels) and $B$ (with $b=3$ levels) in which we wish to test the null hypotheses of no main effect of $A$ by considering the matrix $\mathbf{H}=\mathbf{H}_{A}$. For this scenario, we chose two sample size settings $\mathbf{n}_3= (n_{11}, n_{12},\ldots,n_{23}) = (10,\ldots,10)$ (\textit{balanced}) and $\mathbf{n}_4=(8,10,12,8,10,12)$ (unbalanced), as well as three different censoring scenarios given by the censoring rate vectors $\mathbf{cr}_4=(0.15,0.2,0.25,0.15,0.2,0.25)$ (\textit{medium censoring I}), $\mathbf{cr}_5=(0.3,0.25,0.2,0.3,0.25,0.2)$ (\textit{medium censoring II})
	and $\mathbf{cr}_6 = (0.05,0.1,0.15,0.2,0.15,0.1)$ (\textit{low censoring}).
\end{itemize}
{\bf Distributional choices.} The group-wise survival times $T_{i}$, $i=(i_A,i_B)$, were simulated according to one of the following distributions
\begin{enumerate}
	\item a standard exponential distribution (\textit{Exp}),
	\item a Weibull distribution with  parameters $\lambda_{\text{shape}} = 2$ and $\lambda_{\text{scale}} = \log(2)^{-1/2}$ (\textit{Weib}), 
	\item a standard log-normal distribution (\textit{LogN}),
	\item  different mixture settings (\textit{mix}).
\end{enumerate}
The first three distributions were used for 
the first three sample groups in both factorial designs (a) and (b). In the $2\times2$ layout (a), the fourth group was generated according to a mixture of all three other distributions.
However, in the $2\times3$ layout (b), the fourth, fifth, and sixth groups are pairwise mixtures of the distributions 1. and 3., 1. and 2., 2. and 3., respectively. For the censoring times, we considered uniform distributions $\text{Unif}[0,U_i]$ where the endpoint $U_i$ in group $i$ was calibrated by a Monte-Carlo simulation such that the average censoring rate equals the pre-chosen $cr_i$. Therefore, the formula $cr_i=\P( T_{i1} > C_{i1}) = -\int_0^{\infty} \min\{x/U_i,1\} \,\mathrm{ d }S_i(x)$ was used. The settings described above correspond to the null hypotheses and were used to study the tests' type-1 errors in Section~\ref{sec:sim_typ1}. 

To obtain respective {\bf alternatives}, we shifted the survival and censoring distributions of the first group in the $2\times 2$ layout and of the first two groups for the $2\times 3$ layout by $\delta\in\{0.2,0.4,0.6,0.8,1\}$ to the right. The tests' power performances for these shift alternatives are analyzed in Section~\ref{sec:sim_power}.

{\bf Competing methods.} We included the asymptotic Wald-type tests using the one- and two-sided interval variance estimators with $\gamma = 10\%$ as well as the respective permutation counterparts. Due to the partially high censoring settings, we chose the adaptive procedure introduced in Section~\ref{sec:asy_results} for the two-sided interval strategy. For example, in the scenario with exponentially distributed survival times, $\mathbf{n}=\mathbf{n}_4$, and $\mathbf{cr}=\mathbf{cr}_4$, the upper confidence interval limit $\widehat S_i(l_i)$ did not exist for at least one of the groups in $42\%$ of all considered iterations. Thus, the adaptive procedure with $\tilde \gamma$, $\tilde l_i$, and $\tilde u_i$ instead of $\gamma$, $l_i$, and $u_i$, which we use here, reduces the number of such failed iteration runs, while it coincides with the original two-sided interval method whenever $\widehat S_i(l_i)$ can be determined. We compared these four different testing procedures with a quantile regression method for survival data applied to the median \citep{portnoy03}. The evaluation of the latter was done by means of the jackknife method
\citep{portnoy14}, which is the default choice in the R-package \emph{quantreg} \citep{koenker2020}. Within the regression, the factors were treated as nominal variables and represent, for example, different treatments or different (ethnic) origins.
In addition, the interactions of the factors were included to get a fair comparison with the newly developed Wald-type tests.

The simulations were conducted by means of the computing environment {\bf R} \citep{R}, version 3.6.2, generating $N_{\text{sim}}= 5000$ simulation runs and $N_{\text{res}}=1999$ resampling iterations for the permutation and jackknife procedures. In the rare case that any sample group-specific median does not exist, the corresponding simulation run was not included in the analysis. 
This mostly happened (in around $6\%$ of the cases) for the set-up with censoring $\mathbf{cr}=\mathbf{cr}_3$, $\mathbf{n}=\mathbf{n}_2$ and log-normal or exponential distributed survival times. The nominal significance level was set to $\alpha=5\%$.

\begin{table}[!ht]
	\centering
	\begin{tabular}{ll|lllll|lllll|lllll}
		\multicolumn{2}{c|}{Censoring} & &\multicolumn{3}{c}{low}& & &\multicolumn{3}{c}{high}& && \multicolumn{3}{c}{low/high}& \\
		& & \multicolumn{2}{c}{two-sided} & \multicolumn{2}{c}{one-sided} & & \multicolumn{2}{c}{two-sided} & \multicolumn{2}{c}{one-sided} & & \multicolumn{2}{c}{two-sided} & \multicolumn{2}{c}{one-sided} & \\
		Distr & $\mathbf{n}$ & Per & Asy & Per & Asy & QR & Per & Asy & Per & Asy & QR & Per & Asy & Per & Asy & QR  \\
		\hline
		\multicolumn{17}{c}{Null hypothesis of no main effect}\\
		\hline
		Exp & $\mathbf{n}_1$ & \textbf{5.0} & 4.1 & \textbf{5.0} & 13.5 & 4.2 & \textbf{5.5} & 6.3 & \textbf{4.8} & 6.2 & \textbf{5.5} & 6.9 & 	\textbf{5.4} & \textbf{5.1} & 8.5 & \textbf{5.2} \\ 
		& $2\mathbf{n}_1$ & \textbf{5.2} & \textbf{4.5} & \textbf{5.3} & 9.8 & \textbf{5.4} & 5.8 & \textbf{4.8} & \textbf{5.0} & 5.7 & \textbf{4.8} & 6.2 & \textbf{4.7} & \textbf{5.4} & 7.7 & \textbf{5.2} \\ 
		& $\mathbf{n}_2$ & \textbf{5.0} & 3.7 & \textbf{4.8} & 8.2 & \textbf{5.0} & \textbf{5.4} & 6.4 & \textbf{5.0} & \textbf{5.1} & 6.3 & \textbf{5.0} & 4.3 & \textbf{5.5} & 6.8 & 6.4 \\ 
		& 2$\mathbf{n}_2$ & \textbf{4.5} & 4.0 & \textbf{4.7} & 9.8 & 6.5 & \textbf{5.3} & \textbf{4.8} & \textbf{4.9} & 6.1 & 6.7 & 5.7 & \textbf{4.6} & \textbf{5.1} & 7.6 & 6.1 \\ 
		\hline
		Weib & $\mathbf{n}_1$ & \textbf{5.1} & \textbf{4.8} & \textbf{5.5} & 9.0 & \textbf{5.4} & \textbf{5.1} & 3.8 & \textbf{4.9} & 3.6 & \textbf{5.4} & 6.0 & \textbf{4.6} & \textbf{5.1} & \textbf{5.3} & \textbf{5.4} \\ 
		& $2\mathbf{n}_1$ & \textbf{4.9} & \textbf{4.6} & \textbf{5.1} & 6.6 & \textbf{5.4} & \textbf{5.1} & 3.6 & \textbf{5.1} & \textbf{4.5} & \textbf{4.9} & \textbf{5.4} & \textbf{4.4} & \textbf{5.6} & 5.7 & \textbf{4.8} \\ 
		& $\mathbf{n}_2$ & \textbf{5.5} & \textbf{4.7} & \textbf{5.2} & 6.7 & 6.6 & \textbf{5.0} & 4.2 & \textbf{4.5} & 3.7 & \textbf{4.7}  & \textbf{5.3} & 4.2 & \textbf{4.8} & \textbf{4.4} & 5.8 \\ 
		& 2$\mathbf{n}_2$ & \textbf{4.8} & \textbf{4.6} & \textbf{4.6} & 6.4 & 6.6 & \textbf{5.3} & 4.1 & \textbf{5.3} & \textbf{5.0} & 6.4  & 5.9 & \textbf{5.0} & \textbf{5.6} & 6.0 & 6.5 \\ 
		\hline
		LogN & $\mathbf{n}_1$ & \textbf{5.2} & 3.2 & \textbf{5.0} & 13.5 & \textbf{4.7} & \textbf{5.1} & 5.7 & \textbf{5.1} & 6.5 & \textbf{5.2} & 7.6 & \textbf{5.1} & 5.7 & 9.4 & \textbf{4.6} \\ 
		& $2\mathbf{n}_1$ & \textbf{5.1} & 4.0 & \textbf{4.8} & 10.5 & \textbf{4.5} & \textbf{5.5} & 3.8 & \textbf{5.1} & 6.5 & \textbf{4.6} & 6.1 & 4.1 & \textbf{5.3} & 8.2 & \textbf{4.9} \\ 
		& $\mathbf{n}_2$ & 5.9 & 3.5 & \textbf{5.4} & 9.5 & \textbf{5.4} & \textbf{4.9} & \textbf{5.2} & \textbf{4.8} & 5.8 & \textbf{5.3}  & \textbf{5.6} & 4.2 & \textbf{5.0} & 7.0 & \textbf{4.9} \\ 
		& 2$\mathbf{n}_2$ & \textbf{4.7} & 3.4 & \textbf{5.3} & 10.5 & 6.5 & \textbf{5.3} & 4.3 & \textbf{5.6} & 7.2 & 5.8 & 5.7 & 3.9 & 5.7 & 8.3 & \textbf{5.6} \\ 
		\hline
		mix & $\mathbf{n}_1$ & 5.8 & 4.0 & 6.3 & 12.3 & 5.7 & 6.3 & \textbf{5.5} & 5.8 & 5.7 & 6.1 & 6.6 & 3.8 & 6.2 & 8.2 & \textbf{5.5} \\ 
		& $2\mathbf{n}_1$ & \textbf{4.8} & 4.1 & \textbf{5.2} & 8.6 & 6.1 & \textbf{5.3} & 3.6 & 5.7 & 6.0 & 5.9 & 6.7 & \textbf{4.4} & 6.5 & 7.4 & 5.9 \\ 
		& $\mathbf{n}_2$ & 6.3 & 3.9 & 5.7 & 8.3 & 6.2 & 5.7 & \textbf{5.3} & \textbf{5.3} & \textbf{4.9} & 6.6 & \textbf{5.3} & 3.6 & 6.4 & 7.2 & \textbf{5.4} \\ 
		& 2$\mathbf{n}_2$ & \textbf{5.2} & \textbf{4.5} & 5.8 & 10.4 & 7.2 & \textbf{5.2} & 4.1 & 5.7 & 6.0 & 6.7 & 6.2 & 3.8 & 6.8 & 8.4 & 6.2 \\ 
		\hline
		\multicolumn{17}{c}{Null hypothesis of no interaction effect}\\
		\hline		
		Exp & $\mathbf{n}_1$ & \textbf{4.6} & 3.6 & \textbf{4.7} & 13 & 3.6 & \textbf{5.0} & 5.7 & \textbf{4.5} & \textbf{5.5} & \textbf{4.5} & 7.0 & \textbf{5.4} & \textbf{4.6} & 8.2 & \textbf{4.4} \\ 
		& $2\mathbf{n}_1$ & \textbf{5.3} & \textbf{4.6} & \textbf{5.2} & 10.2 & \textbf{4.4} & \textbf{5.6} & 4.3 & \textbf{5.0} & 6.0 & 4.3 & 6.8 & \textbf{4.9} & 6.0 & 8.0 & \textbf{4.4} \\ 
		& $\mathbf{n}_2$ & \textbf{5.2} & 3.9 & \textbf{4.9} & 8.1 & 4.2 & 4.3 & \textbf{5.6} & \textbf{4.9} & \textbf{5.3} & \textbf{5.0} & \textbf{5.5} & \textbf{4.9} & \textbf{5.4} & 6.5 & \textbf{5.5} \\ 
		& 2$\mathbf{n}_2$ & \textbf{4.8} & \textbf{4.4} & \textbf{4.9} & 10.7 & \textbf{4.9} & \textbf{5.1} & \textbf{4.9} & \textbf{5.3} & 7.0 & \textbf{5.2} & 6.3 & \textbf{5.0} & 5.9 & 9.2 & \textbf{5.5} \\ 
		\hline
		Weib & $\mathbf{n}_1$ & \textbf{5.5} & \textbf{5.0} & \textbf{4.9} & 8.8 & 4.3 & \textbf{4.8} & 3.6 & \textbf{5.4} & 3.8 & 4.2 & 5.8 & 4.0 & \textbf{5.3} & \textbf{5.5} & \textbf{4.6} \\ 
		& $2\mathbf{n}_1$ & \textbf{4.9} & \textbf{4.4} & \textbf{5.2} & 6.5 & 3.9 & \textbf{5.6} & 4.3 & \textbf{5.5} & \textbf{4.8} & 4.2 & \textbf{5.3} & 4.2 & \textbf{5.3} & \textbf{5.5} & \textbf{4.5} \\ 
		& $\mathbf{n}_2$ & \textbf{5.4} & \textbf{4.4} & \textbf{5.5} & 6.6 & \textbf{4.7} & \textbf{4.8} & 4.2 & \textbf{4.9} & \textbf{4.4} & 3.6 & 5.9 & \textbf{4.4} & \textbf{5.1} & \textbf{5.2} & \textbf{4.7} \\ 
		& 2$\mathbf{n}_2$ & \textbf{4.6} & \textbf{4.8} & \textbf{4.7} & 6.8 & \textbf{4.9} & \textbf{5.1} & 4.0 & \textbf{5.5} & \textbf{5.2} & \textbf{4.7} & \textbf{5.3} & \textbf{4.4} & \textbf{5.5} & 6.0 & \textbf{5.0} \\ 
		\hline
		LogN & $\mathbf{n}_1$ & \textbf{5.6} & 3.4 & \textbf{5.4} & 13.8 & 3.9 & \textbf{5.3} & \textbf{5.4} & 5.9 & 7.2 & \textbf{4.6} & 6.9 & 4.3 & \textbf{5.0} & 8.7 & 4.0 \\ 
		& $2\mathbf{n}_1$ & \textbf{4.9} & 3.8 & \textbf{5.0} & 10.0 & 3.8 & \textbf{5.3} & 4.1 & \textbf{5.4} & 6.7 & \textbf{4.4} & 6.4 & 4.2 & \textbf{5.5} & 8.4 & 4.1 \\ 
		& $\mathbf{n}_2$ & \textbf{5.6} & 3.5 & \textbf{5.5} & 9.6 & \textbf{4.5} & \textbf{5.0} & \textbf{5.5} & \textbf{5.3} & 6.7 & \textbf{5.0} & \textbf{5.2} & 3.8 & \textbf{5.1} & 7.5 & 4.1 \\ 
		& 2$\mathbf{n}_2$ & \textbf{4.6} & 3.9 & \textbf{5.2} & 11.5 & \textbf{5.0} & \textbf{5.3} & \textbf{4.5} & \textbf{5.3} & 7.3 & \textbf{4.6} & 6.3 & \textbf{4.4} & 5.8 & 9.4 & \textbf{4.5} \\ 
		\hline
		mix & $\mathbf{n}_1$ & \textbf{5.1} & 3.6 & \textbf{5.6} & 12.2 & 4.0 & \textbf{5.5} & \textbf{4.7} & 5.8 & 5.7 & \textbf{4.6} & 6.8 & 4.0 & 7.0 & 9.2 & 4.1 \\ 
		& $2\mathbf{n}_1$ & \textbf{5.6} & \textbf{4.4} & 6.3 & 9.3 & \textbf{4.9} & \textbf{5.3} & 3.4 & \textbf{5.1} & \textbf{5.2} & 4.2 & 6.7 & 4.2 & 6.3 & 7.6 & \textbf{4.5} \\ 
		& $\mathbf{n}_2$ & 6.0 & 4.0 & \textbf{5.5} & 8.1 & \textbf{4.7} & \textbf{5.5} & \textbf{5.4} & \textbf{5.0} & \textbf{5.0} & \textbf{5.1} & \textbf{5.4} & 3.5 & 5.8 & 6.6 & 4.3 \\ 
		& 2$\mathbf{n}_2$ & \textbf{5.0} & \textbf{4.4} & \textbf{4.9} & 8.9 & \textbf{5.6} & \textbf{4.7} & 3.9 & 5.8 & 6.4 & \textbf{4.6}  & 6.1 & 4.2 & 6.2 & 8.2 & \textbf{4.4} \\ 
		\hline
	\end{tabular}
	\caption{Type-1 error rate in $\%$ (nominal level $\alpha = 5\%$) for
		testing the median null hypotheses of no main effect and no interaction effect, respectively, in the $2\times 2$ design for the asymptotic (Asy) and permutation Wald-type tests bases on the one-sided and two-sided interval variance estimators as well as for the quantile regression method (QR). Values inside the $95\%$ binomial interval $[4.4,5.6]$ are printed bold.}\label{tab:2x2_null}
\end{table}

\begin{table}[ht]
	\centering
	\begin{tabular}{ll|lllll|lllll|lllll}
		\multicolumn{2}{c|}{Censoring} & &\multicolumn{3}{c}{low}& & &\multicolumn{3}{c}{medium I} & && \multicolumn{3}{c}{medium II}& \\
		& & \multicolumn{2}{c}{two-sided} & \multicolumn{2}{c}{one-sided} & & \multicolumn{2}{c}{two-sided} & \multicolumn{2}{c}{one-sided} & & \multicolumn{2}{c}{two-sided} & \multicolumn{2}{c}{one-sided} & \\
		Distr & $\mathbf{n}$ & Per & Asy & Per & Asy & QR & Per & Asy & Per & Asy & QR & Per & Asy & Per & Asy & QR  \\
		\hline
		Exp & $\mathbf{n}_3$  & \textbf{5.0} & 3.5 & \textbf{5.3} & 8.8 & 10.6  & \textbf{5.1} & \textbf{4.6} & \textbf{5.1} & 7.7 & 11.7  & \textbf{5.6} & 2.6 & \textbf{5.1} & 10.6 & 10.1 \\ 
		& $2\mathbf{n}_3$ & \textbf{4.8} & 3.3 & \textbf{4.5} & 6.9 & 9.7 & 5.7 & 3.7 & \textbf{5.4} & 6.6 & 10.4  & \textbf{5.2} & 3.6 & \textbf{4.8} & 8.1 & 9.7 \\ 
		& $\mathbf{n}_4$ & 4.2 & 3.4 & \textbf{4.8} & 8.0 & 10.7 & \textbf{5.2} & \textbf{4.4} & \textbf{5.0} & 7.5 & 12.8 & \textbf{5.5} & 2.8 & \textbf{4.6} & 10.3 & 11.1 \\ 
		& $2\mathbf{n_4}$ & \textbf{5.1} & 3.0 & \textbf{4.9} & 7.1 & 11.4  & 5.8 & 4.1 & \textbf{4.9} & 6.6 & 12.2  & \textbf{5.0} & 3.2 & \textbf{5.3} & 8.5 & 11.7 \\ 
		\hline
		Weib & $\mathbf{n}_3$ & \textbf{5.4} & 3.2 & \textbf{5.4} & 4.3 & 10.4 & \textbf{5.4} & 3.3 & \textbf{5.3} & 3.6 & 11.6 & \textbf{5.1} & 2.9 & \textbf{5.0} & \textbf{5.5} & 11.3 \\ 
		& $2\mathbf{n}_3$ & \textbf{5.1} & 3.6 & \textbf{5.1} & 4.1 & 10.0 &  \textbf{5.2} & 3.4 & \textbf{5.2} & 4.0 & 10.0 & \textbf{4.9} & 3.7 & \textbf{5.2} & \textbf{4.7} & 8.7 \\ 
		& $\mathbf{n}_4$ &  \textbf{4.7} & 2.9 & \textbf{5.6} & \textbf{4.5} & 11.6 & \textbf{5.3} & 3.6 & \textbf{5.4} & 3.8 & 11.1 & \textbf{5.2} & 3.2 & \textbf{5.2} & 5.9 & 10.1 \\ 
		& $2\mathbf{n_4}$ & \textbf{5.0} & 3.4 & \textbf{5.3} & 4.1 & 10.4 & \textbf{5.1} & 3.5 & \textbf{5.3} & 4.0 & 11.4 & \textbf{5.2} & 4.3 & \textbf{5.4} & \textbf{5.2} & 11.5 \\ 
		\hline
		LogN & $\mathbf{n}_3$ & \textbf{5.1} & 3.2 & \textbf{4.8} & 8.4 & 10.0 & \textbf{4.8} & 3.1 & \textbf{5.1} & 7.5 & 10.7 & \textbf{5.3} & 1.9 & \textbf{5.1} & 10.8 & 10.2 \\ 
		& $2\mathbf{n}_3$ & \textbf{5.2} & 2.8 & \textbf{4.7} & 7.1 & 9.3 & \textbf{5.6} & 3.5 & 5.7 & 7.5 & 10.7 & \textbf{5.1} & 2.8 & \textbf{5.1} & 8.7 & 9.5 \\ 
		& $\mathbf{n}_4$ & \textbf{4.5} & 2.6 & \textbf{4.7} & 8.3 & 10.3 & \textbf{5.5} & 4.3 & \textbf{5.0} & 7.6 & 11.2 & \textbf{5.3} & 1.9 & \textbf{5.3} & 10.9 & 11.1 \\ 
		& $2\mathbf{n_4}$ & \textbf{4.9} & 2.8 & \textbf{5.2} & 7.8 & 10.6 & 6.0 & 3.6 & \textbf{4.9} & 6.8 & 11.3 & \textbf{5.4} & 2.9 & \textbf{5.3} & 9.1 & 11.4 \\ 
		\hline
		mix & $\mathbf{n}_3$ & 6.4 & 3.4 & 6.0 & 7.8 & 11.0 & 6.1 & 4.1 & \textbf{5.1} & 5.8 & 12.1 & 6.6 & 2.6 & 5.7 & 9.8 & 11.6 \\ 
		& $2\mathbf{n}_3$ & 6.0 & 3.5 & 6.0 & 6.5 & 10.4 & 6.4 & 3.8 & 6.1 & 6.1 & 11.0 & \textbf{5.1} & 3.2 & 5.7 & 7.3 & 9.6 \\ 
		& $\mathbf{n}_4$ & 6.0 & 3.3 & \textbf{5.6} & 6.9 & 11.7 & 6.0 & 3.6 & \textbf{5.3} & 5.8 & 13.2 & 5.7 & 2.3 & \textbf{4.8} & 8.4 & 11.9 \\ 
		& $2\mathbf{n_4}$ & \textbf{5.2} & 2.9 & 5.7 & 6.4 & 12.3 & 6.9 & 4.3 & 5.9 & 6.1 & 12.7 & \textbf{5.2} & 3.1 & \textbf{5.2} & 7.6 & 11.7
	\end{tabular}
	\caption{Type-1 error rate in $\%$ (nominal level $\alpha = 5\%$) for
		testing the null hypothesis of no main effect of the first factor $A$ in the $2\times 3$ design for the asymptotic (Asy) and permutation Wald-type tests bases on the one-sided and two-sided interval variance estimators as well as for the quantile regression method (QR). Values inside the $95\%$ binomial interval $[4.4,5.6]$ are printed bold.}\label{tab:2x3_null}
\end{table}

\subsection{Type-1 error}\label{sec:sim_typ1}
In this subsection, we analyze the type-1 error rate behavior of all five tests. For an assessment of the results we recall that the standard error of the estimated sizes for the $N=5000$ simulation runs is $0.3\%$ if the true type-1 error probability is indeed $5\%$. This would yield the binomial confidence interval $[4.4\%,5.6\%]$ for the estimated sizes. Values outside this interval deviate significantly from the nominal $5\%$ level.

The simulation results for the $2\times 2$ design are presented in Table~\ref{tab:2x2_null}. It is readily seen that the jackknife approach for the quantile regression method keeps the nominal level quite accurately with a slight tendency to liberal and conservative decisions for tests of no main and no interaction effect, respectively. The two permutation Wald-type test control the nominal level reasonably well, except for the scenarios with simultaneous low and high censoring. Here, the two-sided interval variance estimation leads to rather liberal decisions with values up to $7.6\%$, while the one-sided approach only exhibits a slight liberality under the mixture distribution setting. 
In contrast, the type-1 error probabilities of the asymptotic Wald-type tests strongly depend on the chosen variance estimator: 
on the one hand, the one-sided approach exhibits rather liberal results, which are most pronounced in the low-censoring settings with rejection rates up to $13.8\%$. An exception is the high censoring setting under lognormal distributions; here the decisions are rather conservative with values between $3.6\%$ and $4.4\%$ for the small sample sizes. 
On the other hand, the two-sided strategy leads to rather conservative decisions with values down to $3.2\%$, but the values are frequently (41 out of 96 settings) contained in the binomial confidence interval. 

The type-1 error rates for the $2\times 3$ design are displayed in Table~\ref{tab:2x3_null}. In contrast to the previous observations, the jackknife approach now behaves very liberally with values between $8.7\%$ and $11.9\%$. Both permutation tests keep the type-1 error satisfactorily: under the mixture distribution settings, we can find a slight tendency to liberal decisions with values up to $6.6\%$ and $6.1\%$ for the two- and one-sided variance estimation strategy, respectively. While the asymptotic Wald-type test with the two-sided approach frequently keeps the type-1 error in the $2\times 2$ layout, the decisions are now quite conservative. The conservative behavior is most pronounced in the medium censoring II setting for the lognormal distributions with values  as small as $1.9\%$. 
Comparing this to the one-sided strategy, the decisions again become rather liberal with values up to $10.9\%$ in the medium censoring II setting. However, they come closer to the $5\%$ benchmark when the sample sizes are doubled. An exception to this liberal behavior can be observed under the Weibull distribution, where the decisions are rather conservative with values between $3.6\%$ and $5.5\%$.

In summary, we can generally only recommend the permutation Wald-type tests for both designs and the quantile regression approach for the $2\times 2$ layout. 
All other methods show a quite (partially extremely) liberal or conservative behavior for a significant number of settings. In the next subsection, we analyze how the latter affects the power performances under shift alternatives.

\subsection{Power behavior under shift alternatives}\label{sec:sim_power}

\begin{figure}[ht]
	\centering
	\includegraphics[width=\textwidth]{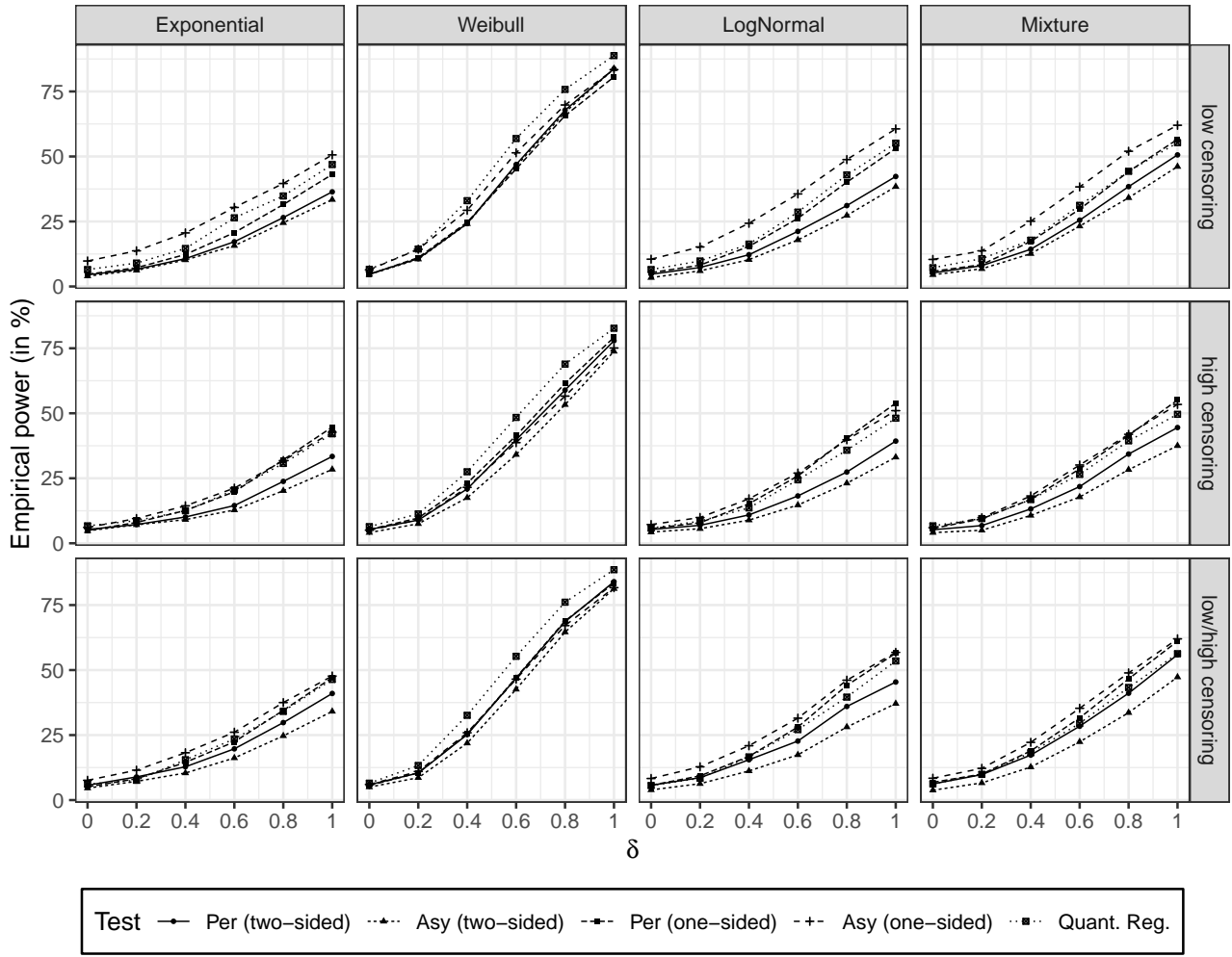}
	\caption{Power curves for the $2\times 2$ design testing for no main effect of the asymptotic (Asy) and permutation (Per) Wald-type tests using the two-sided and one-sided interval variance estimators and of the quantile regression method for $\mathbf{n}=2\mathbf{n}_2$ and shift alternatives $\boldsymbol{\delta}=(\delta,0,0,0)$}\label{fig:2x2_alt_main}
\end{figure}

\begin{figure}[ht]
	\centering
	\includegraphics[width=\textwidth]{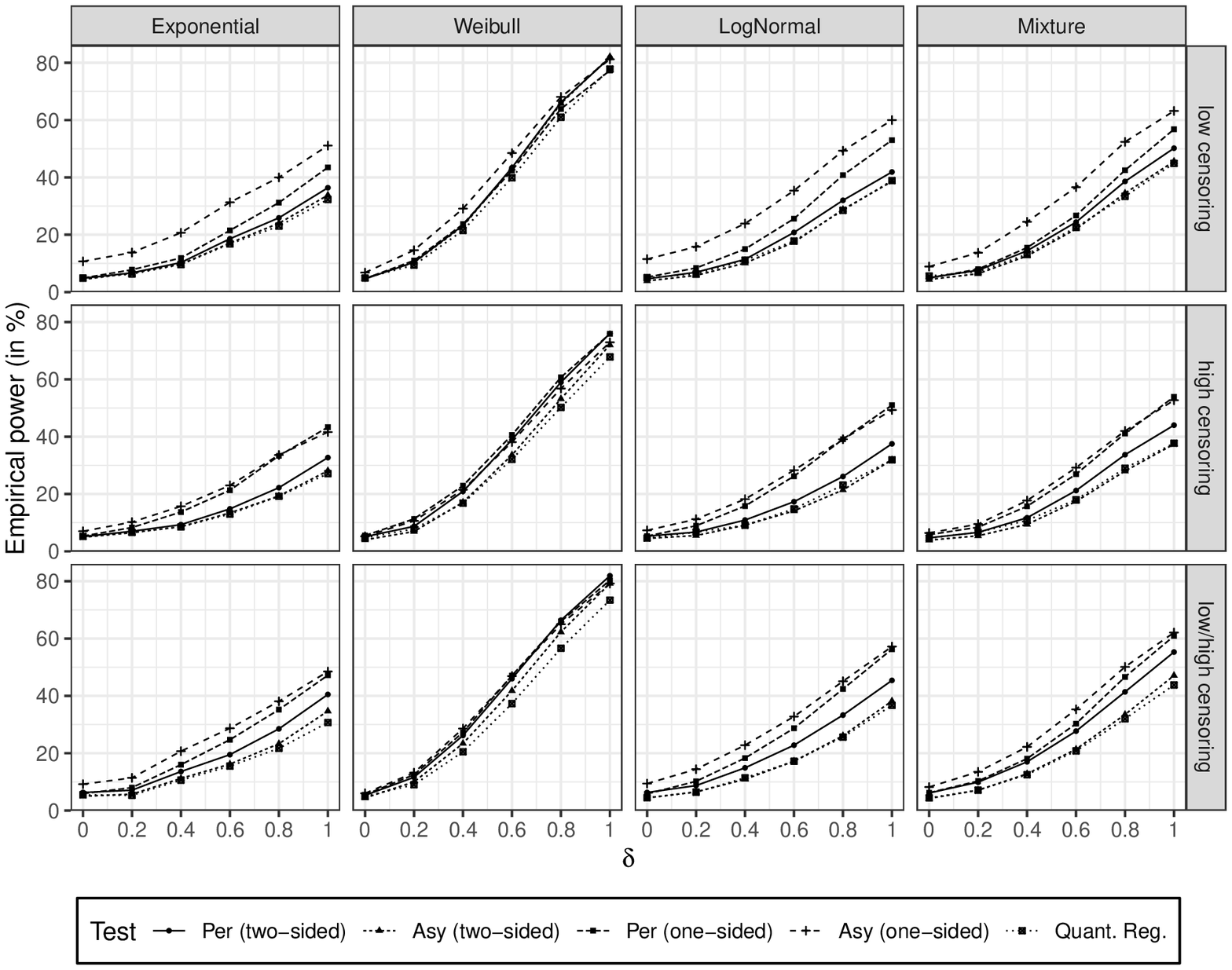}
	\caption{Power curves for the $2\times 2$ design testing for no interaction effect of the asymptotic (Asy) and permutation (Per) Wald-type tests using the two-sided and one-sided interval variance estimators and of the quantile regression method for $\mathbf{n}=2\mathbf{n}_2$ and shift alternatives $\boldsymbol{\delta}=(\delta,0,0,0)$}\label{fig:2x2_alt_inter}
\end{figure}

\begin{figure}[h]
	\centering
	\includegraphics[width=\textwidth]{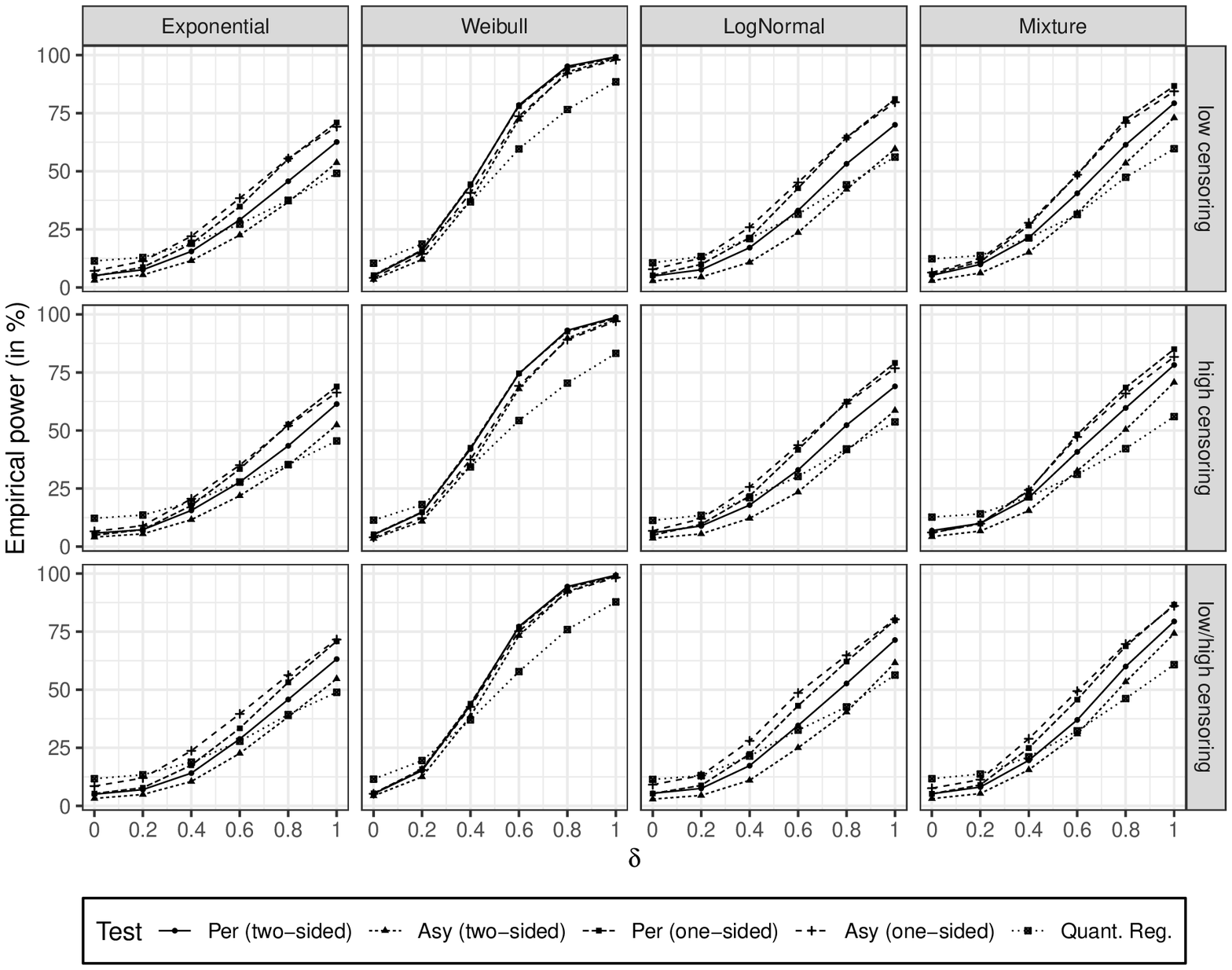}
	\caption{Power curves for the $2\times 3$ design testing for no main effect of the asymptotic (Asy) and permutation (Per) Wald-type tests using the two-sided and one-sided interval variance estimators and of the quantile regression method for $\mathbf{n}=2\mathbf{n}_4$ and shift alternatives $\boldsymbol{\delta}=(\delta,\delta,0,0,0,0)$}\label{fig:2x3_alt_main}
\end{figure}

This subsection is devoted to the comparison of the tests' power performances under shift alternatives, which were introduced at the beginning of this section. We ran the simulations for the larger sample size settings, i.e. $\mathbf{n}=2\mathbf{n}_j$ for $j=1,\ldots,4$. Since the results for the balanced and unbalanced scenarios lead to the same overall conclusions, we only display the results for the respective unbalanced scenarios. 
In Figures~\ref{fig:2x2_alt_main} and \ref{fig:2x2_alt_inter}, the power curves for the $2\times 2$ design testing for no main effect and no interaction effect, respectively, are presented for all five considered tests. Except for testing for no main effect under the Weibull distribution,  the power values of the asymptotic Wald-type test with the one-sided approach show the highest values.

In particular for the low censoring settings the differences to its competitors are most pronounced. This can be explained by the partially very extremely liberal behavior under the null hypotheses, which we observed in the previous subsection. Consequently, the comparison with this specific asymptotic test is not fair at all and need to be taken with a pinch of salt. On the other hand, for the two-sided interval variance strategy, it can be seen that the conservative behavior of the asymptotic tests also affects the power performance: compared to the more accurate permutation approach, they show a significant loss of power. When comparing both permutation tests, it can be seen that the one-sided strategy leads to higher power values in all settings except under Weibull distributions, where the two respective power curves are nearly indistinguishable. 

We finally turn to the most important comparison with the jackknife approach for the quantile regression. A study of Figure~\ref{fig:2x2_alt_main} for the main effect set-up reveals that the curves for the jackknife method is slightly above the ones for the permutation tests under all Weibull settings and for the combination of low censoring and exponentially distributed survival times. However, it should be emphasized that for these four scenarios the observed type-1 error rates of the jackknife method were slightly liberal with values around $6.5\%$ while the permutation tests reliably kept the nominal level. For the remaining cases, the curves of the jackknife method and the permutation tests with the one-sided interval strategy are almost identical. 

These conclusions change completely when testing for interaction effects (Figure~\ref{fig:2x2_alt_inter}). Here, the permutation test with the one-sided interval strategy shows the overall best power behavior while the jackknife quantile regression approach exhibits the lowest power in most situations. 

Finally, the impressions for the $2\times 3$ design (Figure~\ref{fig:2x3_alt_main}) are similar to the $2\times 2$ design for all Wald-type tests. However, a significant difference can be observed for the jackknife method: it now shows much flatter power curves. The latter is rather surprising due to its liberal type-1 error performance under the null hypothesis. 

{\bf Recommendations.} To summarize the simulation results, we recommend the permutation test with the one-sided interval strategy over the other three Wald-type approaches, as it controls the type-1 error most accurately.
It also leads to the highest power values among all Wald-type tests controlling the nominal level $\alpha$. 
A direct comparison of this permutation Wald-type test with the jackknife quantile regression approach clearly favors the Wald-type test when testing for no interaction effect in the $2\times 2$ design and for no main effect in $2\times 3$ layouts. When testing for no main effect in the $2\times 2$ setting both approaches can be recommended.

\section{Illustrative data analysis}\label{sec:real_data}

We illustrate the use of the developed tests by re-analyzing 
a controlled clinical trial conducted in Denmark during the period 1962--1974 \citep{schlichting83}.
Included patients suffered from a histologically verified liver cirrhosis and the times until death have been recorded. 
Due to right-censoring these times were not always observable.
The aim of the study was to analyze the effectiveness of a treatment with prednisone against placebo with respect to survival, while also the influence of several prognostic variables is assessed.
The impact of these additional variables on the survival chances were analyzed by means of a Cox proportional hazards model.
In addition, the authors pointed out on page 892 that ``Another measure for the prognosis calculated from PI (prognostic index, \emph{comment by the authors}) is the median survival time (MST) indicating the span of time that the patient will survive with 50\% probability.''
They also provided a plot of the MST against PI. We complement this descriptive analysis of the MST with inferential investigations based upon our newly developed \emph{nonparametric} methods.

In the following, the dataset \texttt{csl} available through the \texttt{R}-package \emph{timereg} is considered; it contains a subset of size 446 of the original data.
After an initial analysis of the factors treatment and sex, 
we will investigate the influence of the two variables treatment and prothrombin level (at baseline) on the survival median for two specific subsets of the data.
A prothrombin level of less than 70\% of the normal level is considered abnormal \citep[p. 33]{abgk}.

The data sets are analyzed with the help of the five tests, that were compared in the simulation study from Section~\ref{sec:sim}, i.e. the four Wald-type tests developed in this paper (asymptotic/permutation; one-/two-sided interval variance estimator) and the quantile regression method \citep{portnoy03} applied to the median. Ties in the event times have been broken with the help of small random noise. For all tests we choose the significance level $\alpha=5\%$.\\

{\bf Analysis of the complete dataset.} We begin with the analysis of the complete dataset to find out whether there are main and/or interaction effects between the factors treatment and sex on the median survival times.
Based on these outcomes, we will later divide the dataset for a more detailed analysis.
Table~\ref{tab:data_sum1_completeTS} summarizes a few characteristics of the complete dataset.
The sample sizes are quite imbalanced regarding sex; more men than women were included.
Also the censoring rates are higher for the female subgroups which could be caused by a lower mortality rate. The age and prothrombin levels of the individuals at baseline are quite similar which is in line with the fact that patients have been randomized into both treatment groups.

\begin{table}[h]
	\centering
	\begin{tabular}{l|cc|cc|c}
		\multicolumn{1}{r|}{sex}            & \multicolumn{2}{c|}{male} & \multicolumn{2}{c|}{female} &  \\
		\multicolumn{1}{r|}{treatment}       & {placebo} & {prednisone}  & {placebo} & {prednisone}  & overall \\
		\hline
		sample size & 125 & 132 & 95 & 94 & 446 \\
		censoring rate (\%) & 36.0 & 35.6 & 37.9 & 51.1 & 39.5 \\
		av.\ age (years) & 57.9 & 58.6 & 62.1 & 58.9 & 59.2 \\
		av.\ prothrombin level (\% of normal) & 70.8 & 68.5 & 67.1 & 70.2 & 69.2 \\
		est.\ median survival time (years) & 4.43 & 4.37 & 3.20 & 6.74 & 4.78 
	\end{tabular}
	\caption{Summary of the complete dataset according to sex and treatment.}
	\label{tab:data_sum1_completeTS}
\end{table}
\begin{figure}[h]
	\centering
	\begin{minipage}{0.47\textwidth}
		\includegraphics[width=\textwidth]{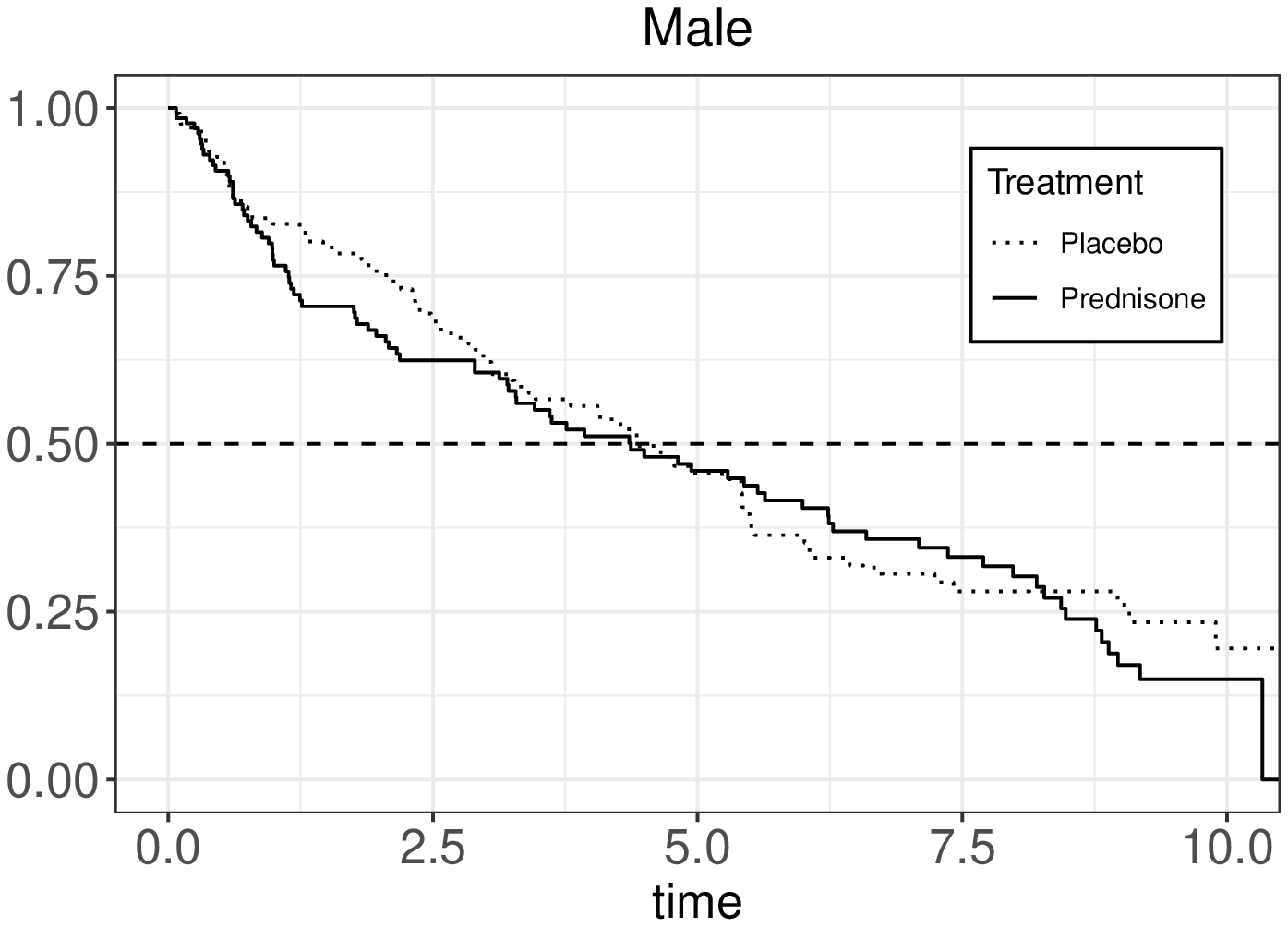}
	\end{minipage}
	\begin{minipage}{0.47\textwidth}
		\includegraphics[width=\textwidth]{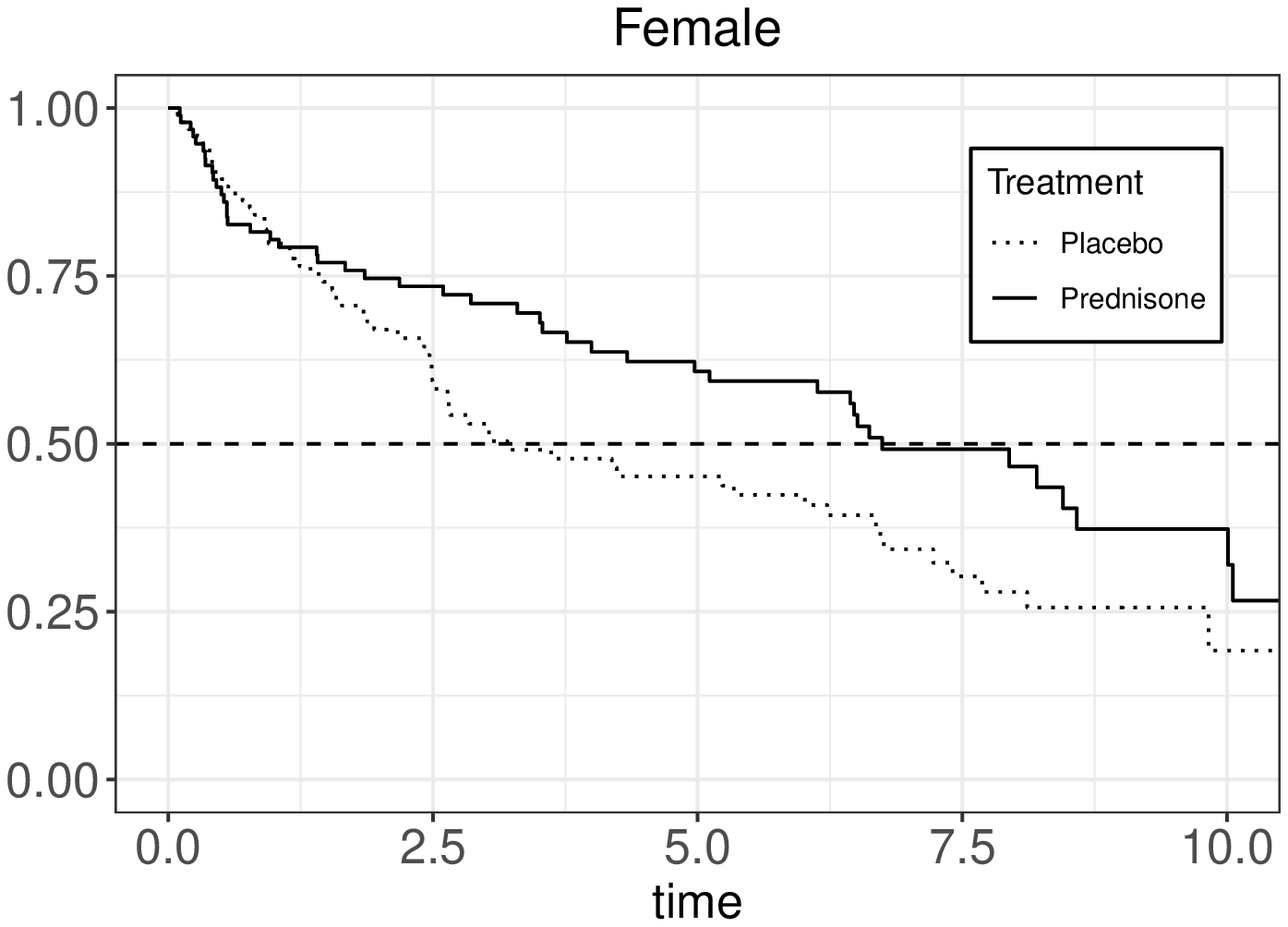}
	\end{minipage}
	\caption{Complete dataset, divided according to sex: male (left panel), female: (right panel). Each panel displays two Kaplan-Meier curves that correspond to both treatments.}\label{fig:data_ex_comp}
\end{figure}
\begin{table}[h]
	\centering
	\begin{tabular}{l|cc|cc|c}
		test procedure & \multicolumn{4}{|c|}{Wald-type test} & QR \\ 
		variance estimation & \multicolumn{2}{|c|}{two-sided} & \multicolumn{2}{|c|}{one-sided} & \\
		approximation method & asymptotic & permutation & asymptotic & permutation & jackknife \\ \hline
		main effect treatment & .060 & .051 & \textbf{.032} & \textbf{.015} & .079 \\
		main effect sex & .527 & .521 & .437 & .425 & .215 \\
		interaction effect & \textbf{.048} & \textbf{.043} & \textbf{.028} & \textbf{.012} & .095
	\end{tabular}
	\caption{$p$-values of all tests applied to the complete dataset with factors treatment and sex. Significant $p$-values (at level $\alpha=5\%$) are printed in bold-type.}
	\label{tab:complete}
\end{table}

The estimated median survival times (Table~\ref{tab:data_sum1_completeTS}) and the plots in Figure~\ref{fig:data_ex_comp} indicate a possible interaction effect between sex and treatment on the MST.
While the medians in both male treatment groups are very close (4.43 and 4.37 years), the medians of the female groups differ quite a lot (3.20 and 6.74 years).
A statistical investigation of the MST shall clarify the question whether there is a significant difference. 
The results of all hypothesis tests can be found in Table~\ref{tab:complete}.
While there is no significant gender effect, only the Wald-type tests based on the one-sided variance estimators found a significant treatment effect. 
The permutation version of this test was the preferred one
according to our simulation study from Section~\ref{sec:sim}. For the tests based on the two-sided variance estimator, we obtain borderline results ($p$-values: .06 and .051).
An interaction effect between treatment and sex has been revealed by all applied Wald-type tests but not by the quantile regression ($p$-value: .095).
Thus, the results of the Wald-type tests are in line with what we expected from our initial numerical and graphical analyses. 
By the way, the original paper \citep{schlichting83} reported a significant sex effect but a non-significant treatment effect within the Cox model. 
However, one should be aware that they the Cox model measures different effects. Moreover, the original study did not consider possible interaction effects and also included additional covariates.\\


%
%
%
%
%
%
%
%

\begin{figure}[ht]
	\centering
	\begin{minipage}{0.47\textwidth}
		\includegraphics[width=\textwidth]{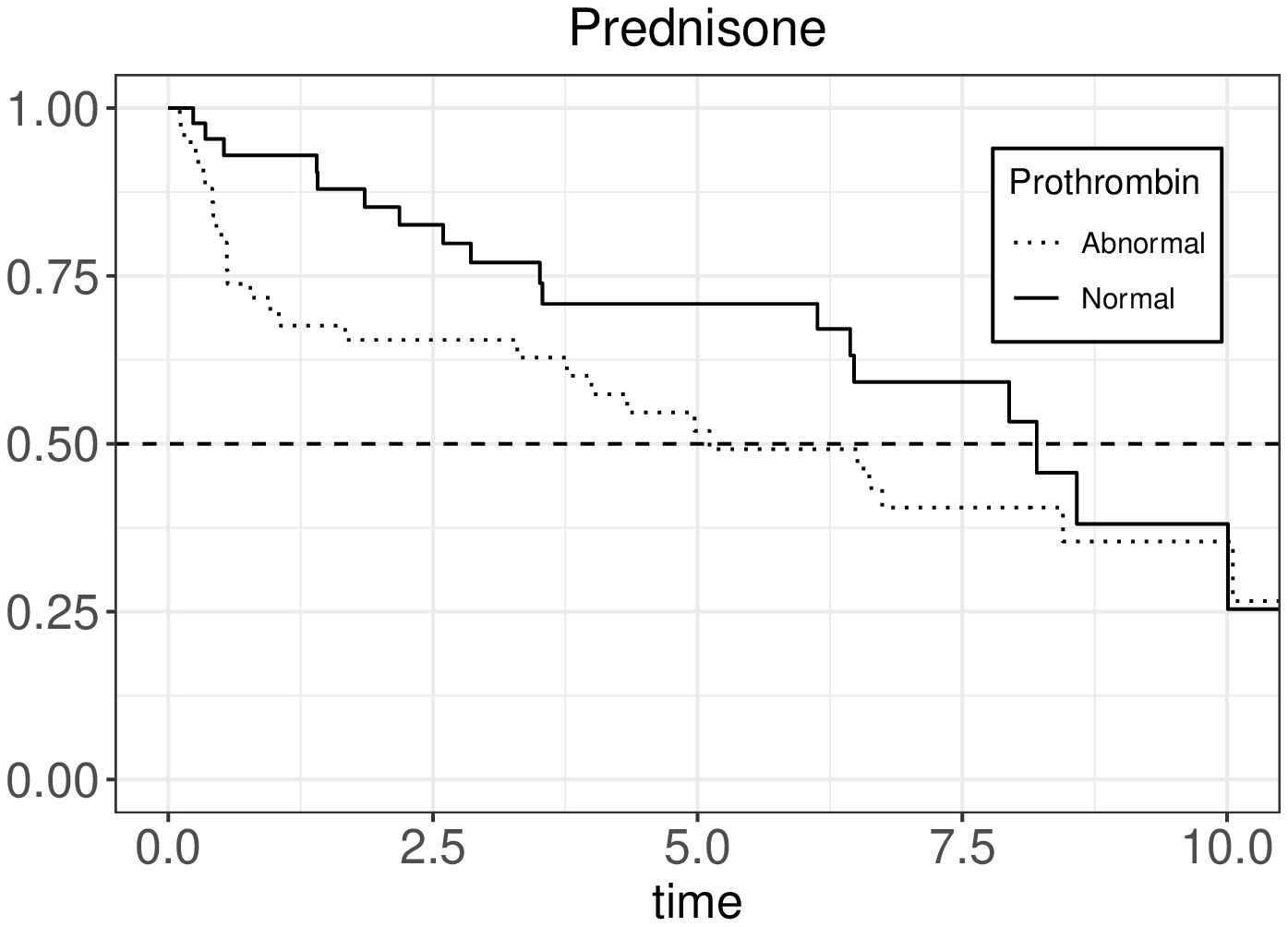}
	\end{minipage}
	\begin{minipage}{0.47\textwidth}
		\includegraphics[width=\textwidth]{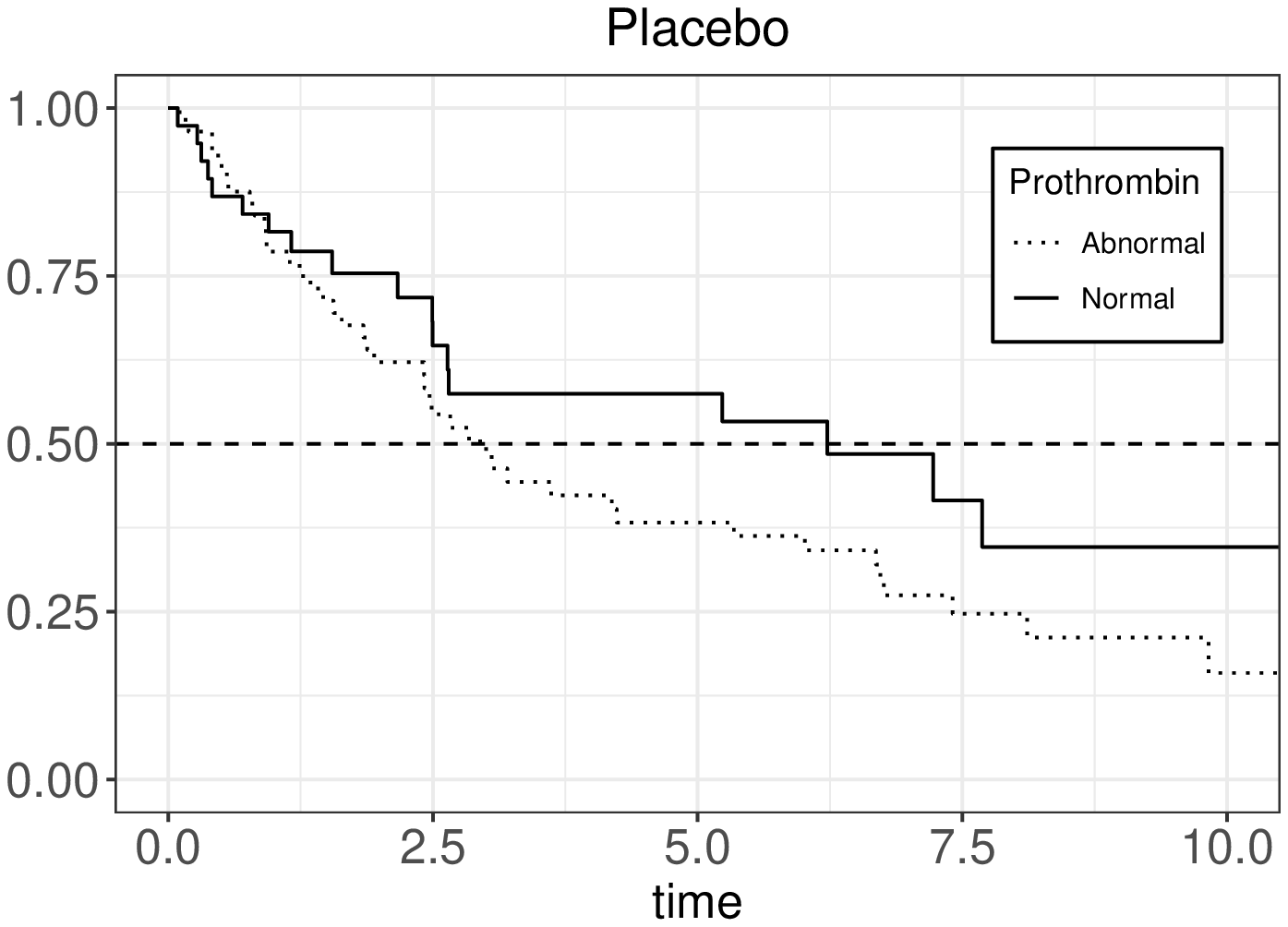}
	\end{minipage}
	\caption{Females subset, divided according to treatment: prednisone (left panel), placebo: (right panel). Each panel displays two Kaplan-Meier curves that correspond to normal/abnormal levels of prothrombin.}
	\label{fig:data_ex_f}
\end{figure}

{\bf Subset analysis of the female group.} Because of the significant influence of sex in the original Cox analysis \citep{schlichting83} and the significant treatment-sex interaction effect in the complete data analysis, we decided to split the dataset according to gender.
At this stage of the analysis, we only focus on the subset of females.
For these we are going to test for main and interaction effects in the factors treatment and normal/abnormal prothrombin level.
A first glance at the summaries in Table~\ref{tab:data_sum3_F_TP} reveal that there are quite big differences in the censoring rates and estimated survival medians depending on normal or abnormal prothrombin levels.
It seems that female patients with normal prothrombin values have a higher chance of survival; see Figure~\ref{fig:data_ex_f}.
This would be in line with the significant protective effect of normal prothrombin levels that were found for the complete dataset in the original Cox analysis \citep{schlichting83}.

\begin{table}[ht]
	\centering
	\begin{tabular}{l|cc|cc|c}
		\multicolumn{1}{r|}{treatment}            & \multicolumn{2}{c|}{placebo} & \multicolumn{2}{c|}{prednisone} &  \\
		\multicolumn{1}{r|}{prothrombin level}       & $<70$ & $\geq 70$ & $<70$ & $\geq 70$ & overall \\
		\hline
		sample size & 57 & 38 & 50 & 44 & 189 \\
		censoring rate (\%) & 28.1 & 52.6 & 44.0 & 59.1 & 44.4 \\
		av.\ age (years) & 62.9 & 61.0 & 60.1 & 57.5 & 60.5 \\
		av.\ prothrombin level (\% of normal) & 50.6 & 91.8 & 50.3 & 92.8 & 68.6 \\
		est.\ median survival time (years) & 3.00 & 6.22 & 5.11 & 8.20 & 6.13
	\end{tabular}
	\caption{Summary of the female subset according to treatment and prothombin level at baseline.}
	\label{tab:data_sum3_F_TP}
\end{table}

\begin{table}[ht]
	\centering
	\begin{tabular}{l|cc|cc|c}
		test procedure & \multicolumn{4}{|c|}{Wald-type test} & QR \\ 
		variance estimation & \multicolumn{2}{|c|}{two-sided} & \multicolumn{2}{|c|}{one-sided} & \\
		approximation method & asymptotic & permutation &asymptotic& permutation & jackknife \\ \hline
		main effect treatment   & .066 & \textbf{.034} & .122 & .136 & .153 \\
		main effect prothrombin & \textbf{.003} & \textbf{.001} & \textbf{.039} & \textbf{.021} & .083 \\
		interaction effect      & .966 & .946 & .972 & .962 & .781
	\end{tabular}
	\caption{$p$-values of all tests applied to the female subset with factors treatment and prothrombin. Significant $p$-values (at level $\alpha=5\%$) are printed in bold-type.}
	\label{tab:female}
\end{table}

All Wald-type tests agree with a significant prothrombin effect. 
The quantile regression, on the other hand, shows a small but still non-significant $p$-value (.083). In addition, most Wald-type tests and also the quantile regression could not detect a treatment effect. This is also in line with the result of the overall Cox analysis         
\citep{schlichting83}.
In addition, none of the tests applied here revealed a treatment-prothrombin interaction effect on the median survival times.

%
%
%

\begin{figure}[ht]
	\centering
	\begin{minipage}{0.47\textwidth}
		\includegraphics[width=\textwidth]{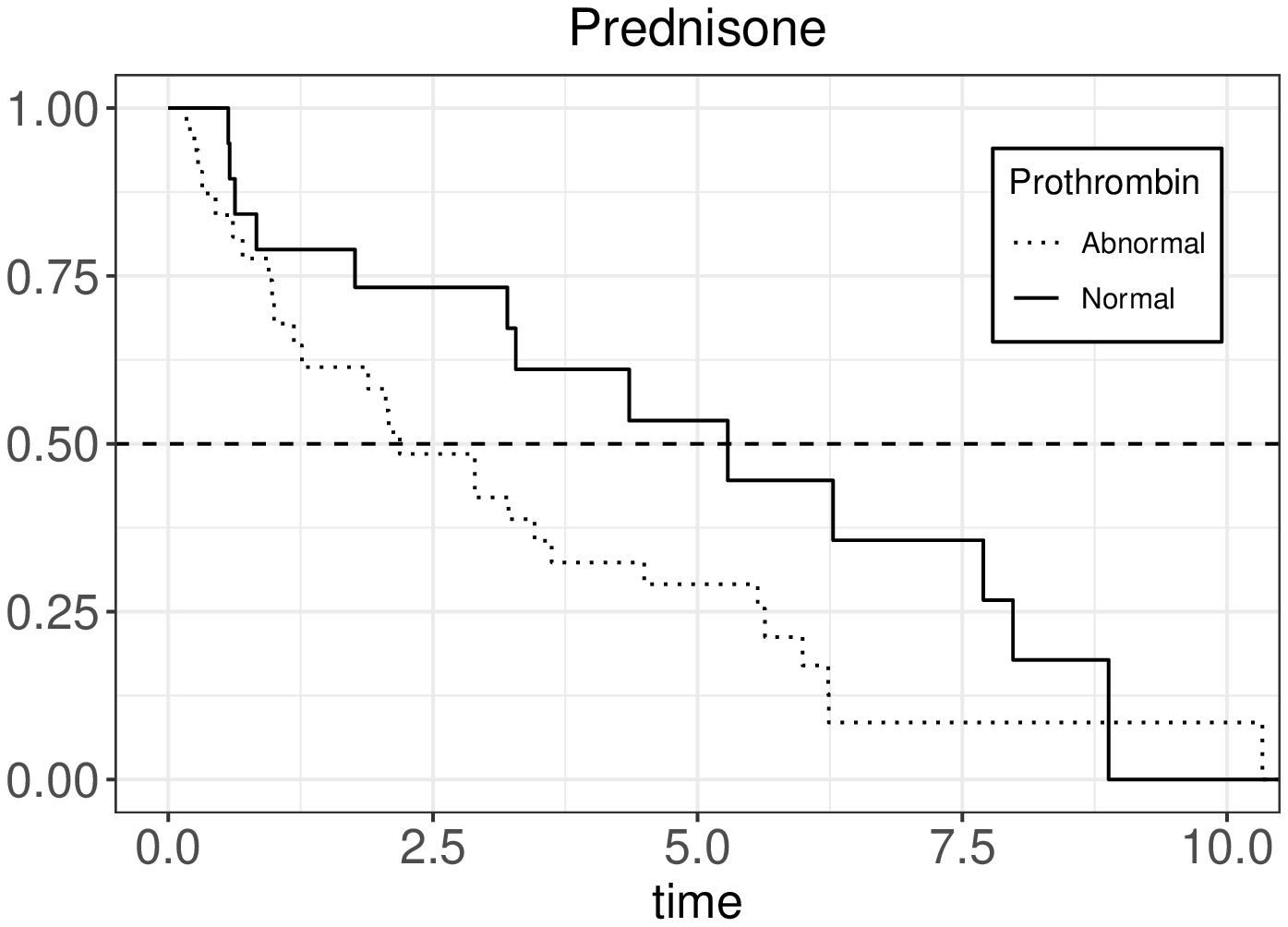}
	\end{minipage}
	\begin{minipage}{0.47\textwidth}
		\includegraphics[width=\textwidth]{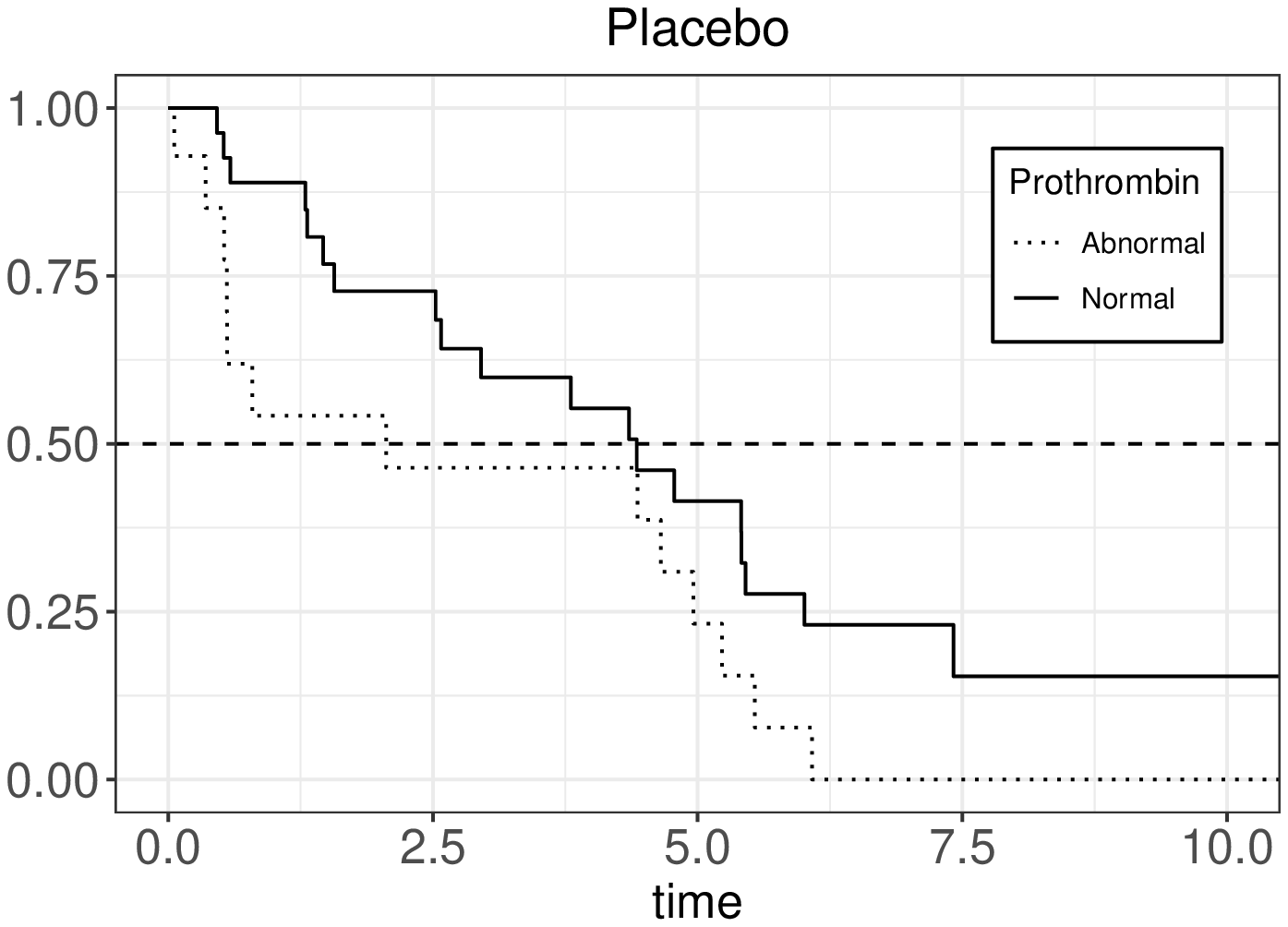}
	\end{minipage}
	
	\caption{Male aged 60--69 subset, divided according to treatment: prednisone (left panel), placebo: (right panel). Each panel displays two Kaplan-Meier curves that correspond to normal/abnormal levels of prothrombin.}
	\label{fig:data_ex_m6069}
\end{figure}

{Subset analysis of males aged 60--69.} In the original Cox analysis \citep{schlichting83}, the variable age was highly significant.
This is why we decided to specifically analyze the subgroup of male patients who were between 60 and 69 years old at baseline. The related Kaplan-Meier curves are displayed in Figure~\ref{fig:data_ex_m6069}. The plots and the summary in Table~\ref{tab:data_sum2_M6069_TP} point out a possible prothrombin effect. 
The sample sizes in all four groups of males divided according to treatment and (ab)normal prothrombin levels are quite small and imbalanced (14, 21, 27, 32);
note that this sample size and censoring setting corresponds to our simulation scenarios with $\mathbf{n}=2\mathbf{n}_2$ and $\mathbf{cr}=\mathbf{cr}_1$ (small differences are due to rounding).
Yet, all applied tests are powerful enough to confirm the presence of a prothrombin effect.
On the other hand, no test found a significant treatment or interaction effect.

\begin{table}[ht]
	\centering
	\begin{tabular}{l|cc|cc|c}
		\multicolumn{1}{r|}{treatment}            & \multicolumn{2}{c|}{placebo} & \multicolumn{2}{c|}{prednisone} &  \\
		\multicolumn{1}{r|}{prothrombin level}       & $<70$ & $\geq 70$ & $<70$ & $\geq 70$ & overall \\
		\hline
		sample size & 14 & 27 & 32 & 21 & 94 \\
		censoring rate (\%) & 7.1 & 29.6 & 12.5 & 38.1 & 22.3 \\
		av.\ age (years) & 65.4 & 64.4 & 63.5 & 63.2 & 64.0 \\
		av.\ prothrombin level (\% of normal) & 49.7 & 87.4 & 53.5 & 89.0 & 70.6 \\
		est.\ median survival time (years) & 2.06 & 4.42 & 2.19 & 5.28 & 3.46
	\end{tabular}
	\caption{Summary of the male aged 60--69 subset according to treatment and prothombin level at baseline.}
	\label{tab:data_sum2_M6069_TP}
\end{table}

\begin{table}[ht]
	\centering
	\begin{tabular}{l|cc|cc|c}
		test procedure & \multicolumn{4}{|c|}{Wald-type test} & QR \\ 
		variance estimation & \multicolumn{2}{|c|}{two-sided} & \multicolumn{2}{|c|}{one-sided} & \\
		approximation method & asymptotic & permutation & asymptotic & permutation & jackknife \\ \hline
		main effect treatment   & .673 & .657 & .624 & .624 & .972 \\
		main effect prothrombin & \textbf{.019} & \textbf{.014} & \textbf{.014} & \textbf{.007} & \textbf{.025} \\
		interaction effect      & .756 & .743 & .714 & .717 & .846
	\end{tabular}
	\caption{$p$-values of all tests applied to the male aged 60--69 subset with factors treatment and prothrombin. Significant $p$-values (at level $\alpha=5\%$) are printed in bold-type.}
	\label{tab:male}
\end{table}

%
%
%
%

\section{Summary and discussion}
\label{sec:discuss}

Factorial designs with time-to-event endpoints are usually analyzed by Cox regression models. 
However, quantification of treatment or interaction effects by a single parameter can only warranted under proportional hazards. This gives rise to study other effect sizes than hazard ratios in factorial designs. In this paper we considered median survival times as they are frequently reported, easy to understand \citep{ben2019median} and have therefore been propagated recently by \cite{chenZhang2016}.

Our main aim was to demonstrate that the analysis of median survival times within general factorial designs is feasible. To this end, we proposed a flexible new class of permutation procedures that can be applied to test 
main and interaction effects or more general contrasts formulated in terms of median survival times. They are based on Wald-type statistics that ensure a robust behavior against non-exchangeable and more complex heterogeneous settings. In fact, we proved consistency of the new methodology under weak conditions; thereby also revealing new insights on permutation empirical processes and variance estimations based upon one-sided confidence intervals which can be found in the technical appendix.

Beyond theoretical investigations, illustrative data analyses and exhaustive simulations for different designs and various distributional settings supported the usage of the new approach. In particular, a permutation test together with a so-called one-sided variance estimation implementation yielded more reliable results than existing methods in our simulation study. 
Moreover, the presented analysis of a real dataset on liver cirrhosis patients confirmed these findings: the proposed permutation Wald-type test is quite powerful and can lead to rejections of null hypotheses even for small sample sizes in combination with moderate censoring rates.


In addition to inference, our tests' outcomes could also be used for factor selection, 
so that more precise inference or confidence regions for (contrasts of) the median survival times of more 
influential factors can be obtained. 

In the future, we plan to transfer the recent findings to ratios of MST contrasts \citep{su1993nonparametric,chen2007covariates}. Moreover, similar extensions of other time-to-event effect sizes, such as the (restricted) mean survival \citep{royston2013restricted, ben2019median} or more general survival quantiles \citep{ditzhaus2019qanova}, to factorial designs will be investigated.
In this context, extensions of the permutation methodology to meta analysis studies based on medians or the restricted mean survival time  \citep{michiels2005meta, wei2015meta}  
will also be part of future research.

To guarantee a simple application of the presented methods, we currently work on the implementation of them into the R-package \textit{GFDsurv} being available on CRAN soon and a corresponding Shiny web application. The respective R-function is coined \textit{medSANOVA}, an abbreviation for median survival analysis-of variance.


\appendix

\section{Proof of Proposition~\ref{PROP:UNCOND}}

As a preparation for the proof regarding the permutation procedure we present a proof of Proposition~\ref{PROP:UNCOND} following the empirical process theory of \cite{vaartWellner1996}, which is different from the proof of \cite{sander:1975}. Fix $i\in\{1,\ldots,k\}$. Let $M>m_i$ be chosen such that $G_i(M)>0$.
The existence of such an $M$ is guaranteed by Assumption~\ref{ass:densities}(b).
Moreover, we introduce the set $\mathbb{D}$ consisting of all non-decreasing and right-continuous functions $H:[0,M]\to \R$ with $H(0)< 0.5$ and $H(M)>0.5$ as well as the set $\mathbb{D}_{b}$ of all bounded functions that are continuous at $m_i$. Both sets are contained in the càdlàg space $D[0,M]$ on the interval $[0,M]$, which we equip with the sup-norm.  Now, let $\Psi:\mathbb{D}\to\R$ be the inverse mapping, compare to Section 3.9.4.2 of \cite{vaartWellner1996}, defined by
\begin{align}\label{eqn:def_Psip}
	\Psi(H) = H^{-1}(0.5) = \inf\{ t\in[0,M] : H(t)\geq 0.5\}.
\end{align}
This functional $\Psi$ is  Hadamard-differentiable \cite[Lemma 3.9.20]{vaartWellner1996} at $H$ tangentially to the space $\mathbb{D}_{b}$ whenever $H$ is differentiable at $m_i$ with positive derivative $h(m_i)>0$. The corresponding Hadamard-derivative is 
\begin{align*}
	\Psi'_{H}(\alpha) = -\frac{ \alpha(m_i)}{h(m_i)}.
\end{align*}
By Example 3.9.31 of \cite{vaartWellner1996},
\begin{align}\label{eqn:process_conv_F_uncon}
	\sqrt{n_i}(\widehat S_{i} - S_i) \overset{d}{\longrightarrow} \mathbb{G}_i \text{ on }D[0,M],
\end{align}
where $\mathbb{G}_i$ is a centred Gaussian process with covariance structure 
\begin{align*}
	(s,t)\mapsto -S_i(t)S_i(s) \int_0^{s\wedge t} \frac{\mathrm{ d }S_i}{G_{i}S_i^2}.
\end{align*}
Consequently, we can conclude from the functional $\delta$-method \citep[Theorem 3.9.4]{vaartWellner1996}  that
\begin{align*}
	\sqrt{n_i}( \widehat S_i^{-1}(1/2) -  S_i^{-1}(1/2) ) = \sqrt{n_i}( \Psi(1-\widehat S_i) - \Psi(1-S_i) ) \overset{ d}{\longrightarrow}  \frac{\mathbb{G}_i(m_i)}{f_i(m_i)}.
\end{align*}
Obviously, the limit is centred normally distributed with variance $\sigma_i^2$ given by \eqref{eqn:def_sigmai}.
$\hfill \qed$

\section{Proof of Lemma \ref{LEM:EST_F_CONSIS}}
In addition to the counting process $Y_i$, we introduce $N_i(t)=\sum_{j=1}^{n_i}\delta_{ij}\mathbf{1}\{X_{ij}\leq t\}$. From the
Glivenko-Cantelli theorem we obtain 
\begin{align}\label{eqn:proof_PB_uni_conv_NY_unc}
	\sup_{t\in[0,M]} \Bigl | \frac{N_i(t)}{n_i} - \nu_i(t) \Bigr| + \sup_{t\in[0,M]} \Bigl | \frac{Y_i(t)}{n_i} - y_i(t) \Bigr| \to 0
\end{align}
almost surely as $n\to\infty$, where $\nu_i(t):= -\int_0^t G_{i}(t) \,\mathrm{ d }S_i(t)$ and $y_i(t):=G_{i-}(t)S_i(t)$, $t\geq 0$, where $G_{i-}$ denotes the left-continuous version of $G_i$. Recall from the proof of Proposition \ref{PROP:UNCOND} that $\widehat m_i \overset{p}{\longrightarrow} m_i$.  Combined with \eqref{eqn:proof_PB_uni_conv_NY_unc} this yields
\begin{align*}
	\widehat V_i = n_i \sum_{j:X_{ij}\leq \widehat m_i} \frac{\delta_{ij}}{Y_i^{2}(X_{ij})} = \int_{0}^{\widehat m_i} \frac{n_i^2}{Y_i^{2}} \,\mathrm{ d }\frac{N_i}{n_i} \overset{p}{\longrightarrow} \int_0^{m_i} \frac{\mathrm{ d }\nu_i}{y_i^2} 
	=: V_i.
\end{align*}
Now, set $\xi_{ni}=0.5z_{\gamma/2} \sqrt{\widehat V_i/n_i}$. By the definition of $\Psi$ in ~\eqref{eqn:def_Psip} we have
\begin{align*}
	\widehat S_{i}^{-1}\Bigl( 0.5 + (-1)^{j+1} \xi_{ni} \Bigr)  = \Psi( 1-\widehat S_{i} + (-1)^{j+1}\xi_{ni}  ).
\end{align*}
Thus, we can rewrite our estimators in terms of $\Psi$ as follows:
\begin{align*}
	\widehat \sigma_{i,\text{two-sided}} &=  0.5  z_{\gamma/2}^{-1} \sqrt{n_i} \Bigl(  \Psi( 1-\widehat S_{i} - \xi_{ni}  ) - \Psi( 1-\widehat S_{i} + \xi_{ni}  )\Bigr), \\
	\widehat \sigma_{i,\text{one-sided}} &=  z_{\gamma/2}^{-1} \sqrt{n_i} \Bigl(  \Psi( 1-\widehat S_{i}) - \Psi( 1-\widehat S_{i} + \xi_{ni}  )\Bigr).
\end{align*}
Since  $\sqrt{n}\xi_{ni}\stackrel p \longrightarrow  0.5z_{\gamma/2} \sqrt{ V_i/\kappa_i}=:\xi_i$, we can conclude from \eqref{eqn:process_conv_F_uncon} that
\begin{align*}
	&\sqrt{n}\Big( 1-\widehat S_{i} - \xi_{ni}  - (1- S_i), 1-\widehat S_{i} + \xi_{in}  - (1- S_i)\Big) \overset{\mathrm d}{\longrightarrow} (-\mathbb{G}_i -\xi_i, -\mathbb{G}_i + \xi_i),\\
	&\sqrt{n}\Big( 1-\widehat S_{i}  - (1- S_i), 1-\widehat S_{i} + \xi_{in}  - (1- S_i)\Big) \overset{\mathrm d}{\longrightarrow} (-\mathbb{G}_i, -\mathbb{G}_i + \xi_i).
\end{align*}
From the Hadamard-differentiability of the inverse mapping $\Psi$, see the explanations below \eqref{eqn:def_Psip}, and the functional $\delta$-method \citep[Theorem 3.9.4]{vaartWellner1996} we can deduce that
\begin{align*}
	&\sqrt{n}\Big( \Psi( 1-\widehat S_{i} -\xi_{ni} ) - \Psi(1-S_i), \Psi( 1-\widehat S_{i} +\xi_{ni} ) - \Psi(1- S_i) \Big)  \overset{ d}{\longrightarrow} \frac{1}{f_i(m_i)}\Bigl( \mathbb{G}_i(m_i) + \xi_i, \mathbb{G}_i(m_i) -\xi_i \Bigr),\\
	&\sqrt{n}\Big( \Psi( 1-\widehat S_{i} ) - \Psi(1- S_i), \Psi( 1-\widehat S_{i} +\xi_{ni} ) - \Psi(1-S_i) \Big)  \overset{ d}{\longrightarrow} \frac{1}{f_i(m_i)}\Bigl( \mathbb{G}_i(m_i), \mathbb{G}_i(m_i) -\xi_i \Bigr).
\end{align*}
Finally, the following convergences also hold in probability because the limits are deterministic:
\begin{align*}
	&\widehat \sigma_{i,\text{two-sided}} \overset{ d}{\longrightarrow}  0.5  z_{\gamma/2}^{-1} \sqrt{\kappa_i} \frac{1}{f_i(m_i)}\Bigl( (\mathbb{G}_i(m_i) + \xi_i)- (\mathbb{G}_i(m_i) - \xi_i) \Bigr) = 0.5 \frac{\sqrt{V_i}}{f_i(m_i)}= \sigma_i,\\
	&\widehat \sigma_{i,\text{one-sided}} \overset{ d}{\longrightarrow}   z_{\gamma/2}^{-1} \sqrt{\kappa_i} \frac{1}{f_i(m_i)}\Bigl( \mathbb{G}_i(m_i)- (\mathbb{G}_i(m_i) - \xi_i) \Bigr) = 0.5 \frac{\sqrt{V_i}}{f_i(m_i)} = \sigma_i.
	\hfill \qed
\end{align*}

\section{Proof of Theorem~\ref{THEO:TESTSTAT_UNCON}(a)}

We temporarily assume the following sample size condition  for all $i=1, \dots, k$ as $n \to \infty$:
\begin{align}
	\frac{n_i}{n}\to\kappa_i >0. \label{eqn:ni_n_kappai}
\end{align}
A combination with the central limit theorem of Proposition~\ref{PROP:UNCOND} makes clear that, as $n \to \infty$, $\mathbf{T} \cdot \sqrt{n}(\widehat{\textbf{m}} - \textbf{m}) \stackrel d \longrightarrow \textbf{Y}$ which has a centered $k$-variate normal distribution with variances $ \kappa_i^{-1} \sigma_i^2,$ $i=1, \dots, k$, and the remaining covariances vanish.

Obviously, $\mathbf{\Sigma} = \text{diag}(\sigma_1^2,\ldots,\sigma_k^2)$ is a regular matrix and $\text{rank}( \mathbf{T} \mathbf{\Sigma} \mathbf{T}') = \text{rank}(\mathbf{T} \mathbf{\Sigma}^{1/2}) = \text{rank}(\mathbf{T})$.
The same holds for $\widehat{\mathbf{\Sigma}}$ instead of  $\mathbf{\Sigma}$.
Because the ranks of the matrices $\mathbf{T} \widehat{\mathbf{\Sigma}} \mathbf{T}'$ never jump and the limit matrix in probability has the same rank,
the Moore--Penrose inverses converge in probability as well.
That is, as $n \rightarrow \infty$,
$( \mathbf{T} \widehat{\mathbf{\Sigma}} \mathbf{T}' )^+ \stackrel p \longrightarrow ( \mathbf{T} \mathbf{\Sigma} \mathbf{T}' )^+$.

It follows from Slutsky's lemma that under $\mathcal{H}_0(\textbf{T})$,
$$W_n(\textbf{T}) = n( \mathbf{T} \widehat{\mathbf{m}})' ( \mathbf{T} \widehat{\mathbf{\Sigma}} \mathbf{T}' )^{+} \mathbf{T} \widehat{\mathbf{m}} \stackrel{d}{\longrightarrow} Z = \mathbf{Y}' ( \mathbf{T} \mathbf{ \Sigma} \mathbf{T}' )^+ \mathbf{Y} $$ 
as $n \rightarrow \infty$.
$Z$ is chi-squared distributed with $\text{rank}( \mathbf{T} \mathbf{\Sigma} \mathbf{T}') = \text{rank}(\mathbf{T})$ degrees of freedom \citep[Theorem 9.2.2]{rao:mitra:1971}.

This limit distribution is independent of the limit proportions $\kappa_i$ from \eqref{eqn:ni_n_kappai}.
Hence, irrespective of the converging subsequences $n_1/n, \dots, n_k/n$, the same limit distribution is obtained as long as $0<\liminf_{n\to \infty} {n_i}/ n \leq \limsup_{n\to \infty} {n_i}/ n<1$, 
i.e.\ under Assumption~\ref{ass:liminf}.
Consequently, the weak convergence holds irrespective of the behaviour of $n_1, \dots, n_k$ as $n \to \infty$ as long as Assumption~\ref{ass:liminf} holds.
\hfill \qed

\section{Proof of Theorem \ref{THEO:TESTSTAT_UNCON}(b)}

We prove the convergence to infinity by showing that $n^{-1}W_n(\mathbf{T}) \stackrel p \longrightarrow ( \mathbf{T} \mathbf{ m})' ( \mathbf{T} \mathbf{  \Sigma} \mathbf{T} )^+ \mathbf{T} \mathbf{m}$ and that the limit is non-zero under $\mathcal{H}_1(\textbf{T}): \mathbf{T}\mathbf{m}\neq \mathbf{0}$.
This alternative hypothesis and the regularity of $\mathbf{\Sigma}^{1/2}=\text{diag}(\sigma_1,\ldots,\sigma_k)$ imply 
that there exists a non-zero vector $\textbf{v} \in \R^k$ such that $\mathbf{m}=\mathbf{\Sigma}^{1/2}\mathbf{v}$.

We use the following properties of Moore--Penrose inverses \citep{rao:mitra:1971}:
for a quadratic matrix $\mathbf{A}$, \  $(\mathbf{A}')^+ = (\mathbf{A}^+)'$, $(\mathbf{A}'\mathbf{A})^+=\mathbf{A}^+(\mathbf{A}')^+$ and $\mathbf{A}\mathbf{A}^+\mathbf{A} = \mathbf{A}$.
From this it follows that
\begin{align*}
	\mathbf{0} \neq \mathbf{T}\mathbf{m} = \mathbf{T}\mathbf{ \Sigma}^{1/2}\mathbf{v} = \mathbf{T}\mathbf{\Sigma}^{1/2}  (\mathbf{T}\mathbf{ \Sigma}^{1/2})^+\mathbf{T} \mathbf{\Sigma}^{1/2}\mathbf{v} 
	= \mathbf{T}\mathbf{ \Sigma}^{1/2} \Bigl[ (\mathbf{T}\mathbf{ \Sigma}^{1/2})^+\mathbf{T}\mathbf{m} \Bigr].
\end{align*}
Hence, $(\mathbf{T} \mathbf{\Sigma}^{1/2})^+ \mathbf{T} \mathbf{m}$ is non-zero.
Finally, 
\begin{align*}
	( \mathbf{T} \mathbf{m})' ( \mathbf{T} \mathbf{  \Sigma} \mathbf{T}' )^+ \mathbf{T} \mathbf{m} = ( \mathbf{T} \mathbf{ m})' ( \mathbf{  \Sigma}^{1/2} \mathbf{T}' )^+ ( \mathbf{T} \mathbf{  \Sigma}^{1/2} )^+ \mathbf{T} \mathbf{m} = \Bigl[ (\mathbf{T}\mathbf{ \Sigma}^{1/2})^+\mathbf{T}\mathbf{m} \Bigr]'\Bigl[ (\mathbf{T}\mathbf{ \Sigma}^{1/2})^+\mathbf{T}\mathbf{m} \Bigr] >0.
	\hfill \qed
\end{align*}

\section{Proof of Theorem \ref{THEO:STAT_PERM}}	
\label{sec:theo_perm}
Analogously to the proof of Theorem \ref{THEO:TESTSTAT_UNCON}, it is sufficient to give the proof for converging sample size proportions $n_i/n$, i.e. under  \eqref{eqn:ni_n_kappai}. Let $\widehat m = \widehat S^{-1} (0.5)$, where $\widehat S$ is the pooled Kaplan--Meier estimator, i.e.  
\begin{align*}
	\widehat S(t)=\prod_{i:X_{ij}\leq t}\Bigl( 1- \frac{\delta_{ij}}{Y(X_{ij})} \Bigr), \quad Y(t)=\sum_{i=1}^kY_i(t),\quad t\geq 0.
\end{align*}
This estimator converges in probability to $S$ defined by
\begin{align*}
	S(t) := \exp\Bigl\{ - \int_0^t \frac{\mathrm{d}\nu}{y} \Bigr\},\quad \mathrm{d}\nu(t):= -\sum_{i=1}^k\kappa_i  G_{i}(t) \,\mathrm{ d }S_i(t), \quad y(t):=\sum_{i=1}^k\kappa_iG_{i-}(t)S_i(t), \quad t\geq 0.
\end{align*}
Let $M >m  $ be such that $\min_{i=1,\dots, k}G_i(M) >0$ which exists because of Assumption~\ref{ass:pooled}.
The pooled Kaplan--Meier estimator even obeys a central limit theorem; see Lemma 2 in the supplement to \cite{dopa2018} for the two-sample case ($k=2$).
An extension to the $k$-sample case, $k\geq 3$, is straightforward: as $n \to \infty$,
\begin{align}
	\sqrt{n}(\widehat S - S) \overset{d}{\longrightarrow} \mathbb{U} \text{ on }D[0,M]  \label{eqn:ctl_S_pooled}
\end{align}	 
for a centred Gaussian process $\mathbb{U}$ with covariance structure 
\begin{align*}
	\E(\mathbb{U}(s)\mathbb{U}(t)) = S(t)S(s) \int_0^{s\wedge t} \frac{\mathrm{ d }\nu}{y^2},\quad s,t\in[0,M].
\end{align*}
It is easy to check  that
\begin{align*}
	f(t):= - \frac{d}{dt} S(t) = \Bigl( -\sum_{i=1}^k\frac{\kappa_i G_{i-}(t)}{y(t)} f_i(t)  \Bigr) \exp\Bigl\{  \sum_{i=1}^k  \int_0^t \frac{\kappa_i G_{i}(s)}{y(s)} \,\mathrm{ d }S_i(s) \Bigr\}, \quad t\geq 0.
\end{align*}
Assumptions~\ref{ass:pooled}(b) and (c) ensure that $f$ is positive and continuous on a neighborhood of $m$. 
The continuous mapping theorem 
yields that $\widehat m = \Psi(1-\widehat S)\overset{p}{\longrightarrow} S^{-1}(0.5) = m$ as $n \to \infty$.
Finally, using similar arguments as for the proof of Proposition~\ref{PROP:UNCOND},  we can prove the asymptotic normality of the permutation median vector. \begin{lemma}\label{lem:perm}
	Suppose \eqref{eqn:ni_n_kappai}. As $n\to\infty$, we have given the data in probability that
	\begin{align}\label{eqn:med_perm}
		\sqrt{n}( \widehat m_1^\pi -  \widehat m,\ldots,\widehat m_k^\pi -  \widehat m)' \overset{ d}{\longrightarrow} \mathbf{Z}^\pi,
	\end{align}
	where $\mathbf{Z}^\pi$ is centred, multivariate normal distributed with covariance matrix $\mathbf{\Sigma}^\pi$ given by its entries
	\begin{align*}
		&\mathbf{\Sigma}^\pi_{ii'} =   \Bigl( \frac{1}{\kappa_i}\mathbf{1}\{ i = i'\} -1 \Bigr)
		0.25 f(m)^{-2} \int_0^{m} \frac{\mathrm{ d }\nu}{y^2}, \quad i,i'=1,\ldots,k.
	\end{align*} 
\end{lemma}  
\noindent The actual proof of Lemma~\ref{lem:perm} is deferred to Appendix~\ref{sec:proof:lem:perm} below.
The covariance matrix can be rewritten as
\begin{align*}
	\mathbf{\Sigma}^\pi = \widetilde{\mathbf{\Sigma}}^\pi - \sigma^{\pi2}\mathbf{J}_k , \quad \widetilde{\mathbf{\Sigma}}^\pi = \text{diag}\Big(\kappa_1^{-1} \sigma^{\pi 2}, \ldots, \kappa_k^{-1} \sigma^{\pi 2} \Big), \quad \sigma^{\pi 2} =  0.25 f(m)^{-2} \int_0^{m} \frac{\mathrm{ d }\nu}{y^2} .
\end{align*}
Since $\mathbf{T} \mathbf{J}_k = \mathbf{0}$, we have $\mathbf{T}\mathbf{\Sigma}^\pi \mathbf{T}'=\mathbf{T}\widetilde{\mathbf{\Sigma}}^\pi \mathbf{T}'$. 
Thus, we can equivalently use in our test statistic an estimator for $\widetilde{\mathbf{\Sigma}}^\pi$, namely the permutation counterpart $\widehat{\mathbf{\Sigma}}^\pi$ of $\widehat{\mathbf{\Sigma}}$, instead of an estimate for the actual limit variance $\mathbf{\Sigma}^\pi$.
It thus remains to show that the permutation version of the interval-based variance estimators are consistent:
\begin{lemma}\label{lem:est_f_perm_consis}
	Suppose  \eqref{eqn:ni_n_kappai}. As $n \to \infty$,
	$\widehat \sigma_{i,\text{one-sided}}^\pi \overset{p}{\longrightarrow} \sigma^\pi$ and $\widehat \sigma_{i,\text{two-sided}}^\pi \overset{p}{\longrightarrow} \sigma^\pi$.
\end{lemma}
\noindent The proof is deferred to Appendix~\ref{sec:proof_lem_est_f_perm_consis} below.
To correctly grasp Lemma \ref{lem:est_f_perm_consis}, we want to point out that (unconditional) convergence in probability is equivalent to conditional convergence in probability given the data. 
Finally, a combination of both lemmas, the continuous mapping theorem, and Theorem 9.2.2~of \cite{rao:mitra:1971} proves Theorem \ref{THEO:STAT_PERM}; compare to the argumentation in the proof of Theorem~\ref{THEO:TESTSTAT_UNCON}.
$\hfill \qed$

\subsection{Proof of Lemma \ref{lem:perm}}\label{sec:proof:lem:perm}

In principle, the statement follows from the argumentation in the proof of Proposition \ref{PROP:UNCOND}. But two aspects need more clarification: first, the joint convergence of the permuted Kaplan--Meier estimators, i.e.\ the (multivariate) permutation version of \eqref{eqn:process_conv_F_uncon}. 
Second, the \emph{uniform} Hadamard-differentiability of the inverse mapping $\Psi$ is required for a permutation variant of the functional delta-method. The latter point was already answered positively by Lemma 8 of~\cite{ditzhaus2019qanova} under some additional assumptions on the location $H$ of differentiation: 
\begin{prop}[Uniform Hadamard differentiability]\label{prop:hada_diff}
	Let $H_n$, $H:[0,M]\to\R$ be nondecreasing, real-valued functions. Moreover, let $H$ be continuously differentiable at $\widetilde m = H^{-1}(1/2)$ with positive derivative $h(\widetilde m)>0$. Suppose that for some $K>0$ 
	\begin{align}
		&\sqrt{n}\sup_{ x \in [0,M]} | H_n(x)-H(x)| \leq K   \text{ for all }n\in\N \label{eqn:lem:hada:cond_Gn-G} \\
		\text{and}\quad &\sqrt{n}\sup_{|x|\leq an^{-1/2}} \Bigl |  H_n(\widetilde m+x) -  H_n(\widetilde m) - H(\widetilde m+x) + H(\widetilde m)\Bigr| \to 0 \text{ for every }a>0. \label{eqn:lem:hada:cond_Gn-Gn-G+G}
	\end{align}
	Then
	\begin{align}\label{eqn:lem:hada_result}
		\sqrt{n} \Bigl( \Psi( H_n +\alpha_n/\sqrt{n}) - \Psi(H_n)  \Bigr) \to \Psi'_{H}(\alpha) = - \frac{\alpha(q)}{g(q)}
	\end{align}
	for every uniformly 
	converging sequence $\alpha_n \to \alpha\in \mathbb{D}_{b,\widetilde m}$ such that $H_n + \alpha_n/\sqrt{n}\in \mathbb{D}$,
	where $\mathbb{D}$ is the set consisting of all non-decreasing and right-continuous functions $H_0:[0,M]\to \R$ with $H_0(0)< 0.5$ and $H_0(M)>0.5$ as well as the set $\mathbb{D}_{b,\tilde m}$ of all bounded functions that are continuous at $\tilde m$. 
\end{prop}
\noindent In the two-sample setting ($k=2$), the joint convergence $\sqrt{n}(\widehat S_1^\pi - \widehat S,\widehat S_2^\pi - \widehat S)$ can be deduced from Theorems~3.7.1 and~3.7.2 of \cite{vaartWellner1996};
cf. Theorem~5 in the supplement to \cite{dopa2018}.
For the general case $k\geq 3$, an extension of these theorems in \cite{vaartWellner1996} was proven in \cite{ditzhaus2019qanova}, see their Lemma 9, and can be applied in the same way as was done for verifying their Lemma 7 to obtain:
\begin{prop}\label{prop:emp_proc_F}
	Under \eqref{eqn:ni_n_kappai}, we have given the data in probability that, as $n\to\infty$,
	\begin{align}\label{eqn:lem:emp_proc_F:result}
		n^{1/2} \Bigl( \widehat S^\pi_{1} - \widehat S,\ldots,\widehat S^\pi_{k} - \widehat S \Bigr) \overset{ d}{\longrightarrow} \mathbb{G}^\pi\quad \text{on }(D[0,M])^k, 
	\end{align}
	where $\mathbb{G}^\pi=(\mathbb{G}^\pi_1,\ldots,\mathbb{G}^\pi_k)$ is a zero-mean Gaussian process on $(D[0,M])^k$ with covariance functions given by 
	\begin{align*}
		\E(\mathbb{G}^\pi_i(s) \mathbb{G}^\pi_{i'}(t)) = \Big(\frac{1}{\kappa_i}\mathbf{1}\{i=i'\} - 1\Big)  S(t)S(s) \int_{0}^{s\wedge t} \frac{\mathrm{ d } \nu}{y^2} 
		,\quad i,i'=1,\ldots,k,\quad s,t\in[0,M].
	\end{align*}
\end{prop}
The convergence in \eqref{eqn:ctl_S_pooled} implies that
\begin{align}\label{eqn:serf2}
	\sqrt{n}\sup_{t\in[0,M]}|\widehat S(t) - S(t)| \stackrel{\ }{\longrightarrow} \sup_{t\in[0,M]}| \mathbb{U}(t)|
\end{align}
in distribution as ${n\to \infty}$.
To apply the functional delta-method for fixed data, we change the underlying probability space to obtain almost surely  the distributional convergence in \eqref{eqn:lem:emp_proc_F:result} and also for the conditions of Proposition~\ref{prop:hada_diff} with $H=1 - S$, $\widetilde m = m=S^{-1}(1/2)$ and $H_n= 1- \widehat S_n$. Since the realization  $\sup_{t\in[0,M]}| \mathbb{U}(t)|(\omega)$ is finite for every fixed event $\omega\in\Omega$, the convergence in \eqref{eqn:serf2} ensures the pointwise boundedness in  \eqref{eqn:lem:hada:cond_Gn-G} on the other space.
It remains to show that \eqref{eqn:lem:hada:cond_Gn-Gn-G+G} holds almost surely on the original probability space which, clearly, is transferable to the other space. 
To change the probability space, we will apply Theorem 1.10.4 of \cite{vaartWellner1996}.

For $a_n=(\log n)^{1/4}/\sqrt{n}$, it follows from Theorem 1 of \cite{cheng:1984} that we have almost surely
\begin{align*}
	\sup_{ |x|\leq a_n} \Bigl | \widehat S_{i}(m_i+x) - \widehat S_{i}(m_i) - S_i(m_i+x) + S_i(m_i)\Bigr| = o(n^{-2/3})
\end{align*}
for every $i \in \{1,\ldots,k\}$ if $f_i$ is continuous on a neighborhood of $m_i$.
Instead of this result, in order to show~\eqref{eqn:lem:hada:cond_Gn-Gn-G+G}, we need an analogue for the pooled Kaplan--Meier estimator around the pooled median:
\begin{align}\label{eqn:conv_S_pooled}
	\sup_{ |x|\leq a_n} \Bigl | \widehat S(m+x) - \widehat S(m) - S(m+x) + S(m)\Bigr| = o(n^{-2/3}).
\end{align}
This can be shown by slightly adapting the proof of Theorem~1 in \cite{cheng:1984}.
In particular, one should first note that all algebraic manipulations that are made in the just mentioned proof can be applied in the same way to the pooled quantities.
Next, one can find a decomposition of the pooled Kaplan-Meier estimator into simpler functions, similarly as in display~(2.2) of  \cite{cheng:1984}:
\begin{align*}
	1- \widehat F(t) = \exp \Big( \sum_{s \in D_n \cap [0,t]} \log \frac{\sum_{i=1}^k Y_i(s+) }{\sum_{i=1}^k Y_i(s)} \Big)
	=\exp \Big( \sum_{s \in D_n \cap [0,t]} \log \frac{ Y(s+) }{ Y(s)}\Big)
\end{align*}
where $D_n$ is the set of discontinuities of the pooled Kaplan-Meier estimator, i.e.\ the set of all uncensored event times, $Y_i$ is the left-continuous number at risk function in group $i$, and $Y$ is the pooled at risk function. The plus sign after the argument $s$ indicates the right-hand limit of a function at $s$.

In order to proceed analogously as in the proof of Theorem~1 in \cite{cheng:1984}, one needs to establish an inequality similar to (2.4) therein, i.e.
\begin{align}
	\label{eq:foeldes}
	P\Big(\sup_{0 \leq t \leq M} | \widehat S(t) - S(t) | > \varepsilon \Big) \leq d_0 \exp(-n \varepsilon^2 \delta^4 d_1)
\end{align}
for certain positive constants $d_0, d_1$.
Such an inequality can be deduced by following the lines of \cite{foeldes81}, 
i.e.\ starting from the decomposition 
\begin{align*}
	\log \widehat S(t) & = \sum_{j=1}^n \delta_j 1\{ X_j \leq u \} \log \frac{Y(X_j +)}{Y(X_j)} \\
	& = - \frac1n \sum_{j=1}^n \delta_j 1\{ X_j \leq u \} y(X_j)^{-1}
	- \frac1n \sum_{j=1}^n \delta_j 1\{ X_j \leq u \} \sum_{\ell=2}^\infty \frac1\ell Y(X_j)^{-\ell} \\
	& \quad - \frac1n \sum_{j=1}^n \delta_j 1\{ X_j \leq u \} (n Y(X_j)^{-1} - y(X_j)^{-1}).
\end{align*}
The last equality is due to a power series expansion of $x \mapsto \log(1 + x) $ around $- 1/Y(X_j) $.
Now it is clear that all three terms on the right-hand side can be treated similarly as in the proofs of \cite{foeldes81}, after the sums over the individuals $j$ have been divided into the sample-specific sums $\sum_{i=1}^k \sum_{j=1}^{n_i}$, 
and \eqref{eq:foeldes} follows.

Now, if one replaces the (sub)survival functions etc.\ with the pooled counterparts, the rest of the proof of Theorem~1 in \cite{cheng:1984} can be paralleled without further difficulties because the remaining algebraic manipulations still apply.
It should just be noted that the utilized large deviation inequalities (by Hoeffding, Bernstein, and also one by Kiefer) are still applicable if one, again, first divides the sums into sample-specific sums: e.g.\ for some $\varepsilon >0$ and some sum $\sum_{j=1}^n f(X_{j})$, which often appears in the mentioned proof, we estimate
$$ P \Big( \frac1n \sum_{j=1}^n f(X_{j}) > \varepsilon \Big) =  P \Big(  \sum_{i=1}^k \frac{n_i}{n} \frac1{n_i} \sum_{j=1}^{n_i} f(X_{ij}) > \varepsilon \Big) 
\leq 
\sum_{i=1}^k P \Big( \frac1{n_i} \sum_{j=1}^{n_i} f(X_{ij}) >  \frac{n}{n_i} \frac{\varepsilon}{k}  \Big).$$
Note that $ \frac{n}{n_i} \frac{\varepsilon}{k} \to \varepsilon/(\kappa_i k) > 0$.  In this way, all utilized large deviation theorems remain applicable and, eventually, the convergence in \eqref{eqn:conv_S_pooled} follows.

Now, we fix the data.
Regarding the argumentation above about changing the probabilty space, we can assume without loss of generality that \eqref{eqn:lem:hada:cond_Gn-G}, \eqref{eqn:lem:hada:cond_Gn-Gn-G+G}, and  \eqref{eqn:lem:emp_proc_F:result} hold.
An application of the (uniform) functional $\delta$-method \citep[Theorem 3.9.5]{vaartWellner1996} with the map $\Gamma: (H_1,\ldots,H_k) \mapsto (\Psi(H_1)  ,\ldots,\Psi(H_k))$ yields
\begin{align*}
	\sqrt{n}\Big( (\widehat S_i^\pi)^{-1}(1/2) -  \widehat S^{-1}(1/2) \Big)_{i=1,\ldots,k} 
	&= \sqrt{n}\Big( \Gamma(1-\widehat S_1^\pi,\ldots,1-\widehat S_k^\pi) 
	- \Gamma(1-\widehat S,\ldots,1-\widehat S) \Big)  \\
	&\overset{\mathrm d}{\longrightarrow}  \frac{\mathbb{G}^\pi(m)}{f(m)} \overset{\mathrm d}{=} \mathbf{Z}^\pi.
\end{align*}
Finally, this translates back to the distributional convergence~\eqref{eqn:med_perm} given the data in probability on the original probability space.
\hfill \qed

\subsection{Proof of Lemma \ref{lem:est_f_perm_consis}}
\label{sec:proof_lem_est_f_perm_consis}
In addition to the counting processes $Y_i$ and $Y$, we introduce $N_i(t)=\sum_{j=1}^{n_i}\delta_{ij}\mathbf{1}\{X_{ij}\leq t\}$ and $N= \sum_{i=1}^kN_i$. From the
Glivenko-Cantelli theorem we obtain 
\begin{align}\label{eqn:proof_PB_uni_conv_NY}
	\sup_{t\in[0,M]} \Bigl | \frac{N(t)}{n} - \nu(t) \Bigr| + \sup_{t\in[0,M]} \Bigl | \frac{Y(t)}{n} - y(t) \Bigr| \to 0 
\end{align}
almost surely as $n\to\infty$.
Recall from the proof of Theorem \ref{THEO:STAT_PERM} that $\widehat m \overset{p}{\longrightarrow} m$. In particular, for every subsequence of increasing sample sizes there exists a further subsequence such that $\widehat m \to m$ almost surely along the latter subsequence. 
Throughout the rest of the proof, we fix the observations. Without a loss of generality and operating along appropriate subsequences, \eqref{eqn:med_perm} holds and we can treat, from now on, $\widehat m$ as a sequence of constants converging to $m$ and $N$ as well as $Y$ as non-random functions fulfilling the uniform convergence in \eqref{eqn:proof_PB_uni_conv_NY}. Let $Y_i^\pi$ and $N_i^\pi$ the permutation counterparts of $Y_i$ and $N_i$. By \cite{neuhaus:1993}, see his equation (6.1),
\begin{align*}
	\sup_{t\in[0,M]} \Bigl | \frac{Y_i^\pi(t)}{Y(t)} - \kappa_i \Bigr| \overset{p}{\longrightarrow} 0.
\end{align*}
Using similar arguments, the statement remains true when one replaces $Y_i^\pi$ and $Y$ by $N_i^\pi$ and $N$, respectively. Combining this with $\widehat m_i^\pi \overset{p}{\longrightarrow} m$, which follows from \eqref{eqn:med_perm} and $\widehat m \to m$, and \eqref{eqn:proof_PB_uni_conv_NY} as well as rewriting $\mathbf{X}_i^\pi=(X_{i1}^\pi, \delta_{i1}^\pi,\ldots,X_{in_k}^\pi, \delta_{in_k }^\pi)$ yields
\begin{align*}
	\widehat V_i^\pi = n_i \sum_{j:X_{ij}^\pi\leq \widehat m_i^\pi} \frac{\delta_{ij}^\pi}{Y_i^{\pi 2}(X_{ij}^\pi)} = \frac{n_i}{n}\int_{0}^{\widehat m_i^\pi} \frac{n^2}{Y_i^{\pi 2}} \,\mathrm{ d }\frac{N_i^\pi}{n} \overset{p}{\longrightarrow} \int_0^{m} \frac{\mathrm{ d }\nu}{y^2} 
	=: V^\pi.
\end{align*}
Now, set $\xi_{ni}^\pi=0.5z_{\gamma/2} \sqrt{\widehat V_i^\pi/n_i}$. Similar to the proof of Lemma~\ref{LEM:EST_F_CONSIS}, we can rewrite the estimators in terms of the inverse mapping $\Psi$, see \eqref{eqn:def_Psip}, as follows:
\begin{align*}
	\widehat \sigma^\pi_{i,\text{two-sided}} &=  0.5  z_{\gamma/2}^{-1} \sqrt{n_i} \Bigl(  \Psi( 1-\widehat S_{i}^\pi - \xi_{ni}^\pi  ) - \Psi( 1-\widehat S_{i}^\pi + \xi_{ni}^\pi  )\Bigr), \\
	\widehat \sigma^\pi_{i,\text{one-sided}} &=  z_{\gamma/2}^{-1} \sqrt{n_i} \Bigl(  \Psi( 1-\widehat S_{i}^\pi) - \Psi( 1-\widehat S_{i}^\pi + \xi_{ni}^\pi  )\Bigr)
\end{align*}
Since  $\sqrt{n}\xi_{ni}^\pi\stackrel p \longrightarrow  0.5z_{\gamma/2} \sqrt{ V_i^\pi/\kappa_i}=:\xi_i^\pi$, we can conclude, in analogy to the proof of Lemma \ref{lem:est_f_perm_consis} and the proof of Lemma~\ref{lem:perm}, from Lemma \ref{prop:hada_diff} and the uniform functional $\delta$-method \citep[Theorem 3.9.5]{vaartWellner1996} that
\begin{align*}
	&\widehat \sigma^\pi_{i,\text{two-sided}} \overset{ d}{\longrightarrow}  0.5  z_{\gamma/2}^{-1} \sqrt{\kappa_i} \frac{1}{f(m)}\Bigl( (\mathbb{G}_i^\pi(m) + \xi_i^\pi)- (\mathbb{G}_i^\pi(m) - \xi_i^\pi) \Bigr) = 0.5 \frac{\sqrt{V^\pi}}{f(m)},\\
	&\widehat \sigma^\pi_{i,\text{one-sided}} \overset{ d}{\longrightarrow}   z_{\gamma/2}^{-1} \sqrt{\kappa_i} \frac{1}{f(m)}\Bigl( \mathbb{G}_i^\pi(m)- (\mathbb{G}_i^\pi(m) - \xi_i^\pi) \Bigr) = 0.5 \frac{\sqrt{V^\pi}}{f(m)}.
\end{align*}

\section*{Acknowledgement}
Marc Ditzhaus and Markus Pauly were supported by the {Deutsche Forschungsgemeinschaft} (Grant no.  PA-2409 5-1). 
The authors thank Stefan Inerle for computational assistance.

\bibliographystyle{chicago}
\bibliography{sample}

\end{document}